\documentclass[journal]{IEEEtran}
\usepackage{placeins}
\usepackage{tabstackengine}
\setstackEOL{\cr}
\usepackage{color}
\usepackage{amsmath}
\usepackage{amssymb}
\usepackage[thmmarks, amsmath]{ntheorem}
\usepackage{wrapfig}
\usepackage{epsfig}
\usepackage{epstopdf}
\usepackage[caption=false,font=footnotesize]{subfig}
\usepackage{multirow}
\usepackage{pdflscape}
\usepackage{makecell}
\usepackage[noadjust]{cite}
\usepackage{nicefrac}
\usepackage{stackengine}
\usepackage[cal=cm]{mathalfa}
\usepackage{float}
\usepackage{stfloats}
\usepackage{graphicx}
\usepackage{mathtools}
\newtheorem{definition}{Definition}

\usepackage[table,xcdraw]{xcolor}

\begin{document}

\title{ 
Small-Signal Stability and Hardware Validation of Dual-Port Grid-Forming Interconnecting Power Converters in Hybrid AC/DC Grids
}

\author{Josep Arévalo-Soler, Mehrdad Nahalparvari, Dominic Gro{\ss}, Eduardo Prieto-Araujo, Staffan Norrga, \\ Oriol Gomis-Bellmunt\thanks{Josep Arévalo Soler, Eduardo Prieto Araujo, and Oriol Gomis Bellmunt are with the CITCEA-UPC, 08028 Barcelona, Spain (e-mail: josep.arevalo@upc.edu; eduardo.prieto-araujo@upc.edu; oriol.gomis@upc.edu).
Mehrdad Nahalparvari and Staffan Norrga are with the KTH Royal Institute of Technology, 10044 Stockholm, Sweden (e-mail: mnah@kth.se; norrga@kth.se).
Dominic Groß is with the Electrical and Computer Engineering, University of Wisconsin-Madison, Madison, WI 53706 USA (e-mail: dominic.gross@wisc.edu).}
}

\maketitle
\date{}

\begin{abstract}
Interconnecting power converters (IPCs) are the main elements enabling the interconnection of multiple high-voltage alternating current (HVAC) and high-voltage direct current (HVDC) subgrids. To ensure stable operation of the resulting hybrid ac/dc systems, grid-following (GFL) and grid-forming (GFM) controls need to be carefully assigned to individual IPC terminals when using common IPC controls. 
In contrast, dual-port GFM control imposes a stable voltage on the ac and dc terminals and can be deployed on all IPCs regardless of the network configuration. In this work, we use hybrid ac/dc admittance models, eigenvalue sensitivities, and case studies to analyze and quantify the underlying properties of AC-GFM control, AC-GFL, and dual-port GFM control. Compared to common AC-GFM and AC-GFL controls, dual-port GFM control (i) renders IPCs dissipative over a much wider range of frequencies and operating points, (ii) significantly reduces the sensitivity of IPC small-signal dynamics to operating point changes, and (iii) exhibits an improved dynamic response to severe contingencies. Finally, the results are illustrated and validated in an experimental scaled-down point-to-point HVDC system.
\end{abstract}


\section{Introduction}
The increasing trend towards integrating renewable energy sources into bulk power systems is introducing new engineering solutions, such as high voltage direct current (HVDC) transmission, that enable the transmission of electric energy across long distances~\cite{Llibre_Oriol}. Well-known examples include the planned energy islands in the North Sea and the Baltic Sea~\cite{energy_island3} that are envisioned to rely on multi-terminal high voltage direct current systems (MT-HVDC)~\cite{North-Sea} and MT-HVDC systems in China~\cite{Nanao, Zhoushan, Zhangbei1, Zhangbei2}. These projects will enable the interconnection of large-scale offshore wind resources to legacy onshore power systems. Moreover, interconnecting existing transmission corridors in the North Sea will result in a hybrid ac/dc system connecting different asynchronous ac grids {through} MT-HVDC systems. This integration is enabled by so-called interconnecting power converters (IPC) that rely on voltage source converter (VSC) technology, more specifically modular multilevel converters (MMCs)~\cite{Llibre_MMC}, to interface between ac and dc subsystems.

In this context, a key challenge is selecting the control strategy for each IPC. Broadly speaking, IPC controls can be classified into grid-following (GFL) or grid-forming (GFM) controls. This classification is usually applied with respect to the ac terminal (i.e., AC-GFL and AC-GFM). 

AC-GFL controls assume a stable ac voltage (i.e., magnitude and frequency) and typically rely on a phase-locked loop (PLL) {to establish network synchronization}~\cite{22}. {While AC-GFL control can provide grid-support functions, stability often rapidly deteriorates under weak grid conditions or} due to insufficient AC-GFM units. In contrast, AC-GFM control {imposes a} stable ac voltage {(i.e., magnitude and frequency)} at the converter terminal. {Various} AC-GFM {controls} such as Virtual Synchronous Machine (VSM) control~\cite{27}, droop control~\cite{pf_1}, dispatchable Virtual Oscillator Control (VOC)~\cite{GFM1} and machine emulation control~\cite{30} {have been proposed, among others}. The {categorization into GFL} and GFM control can also be applied {to} the {IPC} dc terminal ({i.e.,} DC-GFL and DC-GFM)~\cite{dualport}. {Broadly speaking,} DC-GFL controls {require} a stable dc voltage at the {IPC} terminal {while} DC-GFM controls {impose} a stable dc voltage at {the IPC dc} terminal~\cite{dualport}. {Ultimately, VSC controls can be categorized into  four broad classes:} AC-GFL/DC-GFL, AC-GFM/DC-GFL, AC-GFL/DC-GFM, and AC-GFM/DC-GFM. {Out of these categories, only AC-GFM/DC-GFM is able to stabilize both the ac and dc terminal voltage. In contrast, AC-GFL/DC-GFM requires a stable ac voltage, AC-GFL/DC-GFL requires a stable ac and dc voltage at its converter terminals and AC-GFM/DC-GFL requires a stable dc voltage to operate.}

We emphasize that common IPC controls only cover the AC-GFL/DC-GFL, AC-GFM/DC-GFL, and AC-GFL/DC-GFM categories. Notably, controls applied to the MMCs are often inspired by structures employed in two-level VSC and using uncompensated modulation (UCM)~\cite{UCM_3,UCM_jeff}, which  couples the ac and dc terminals. However, using compensated modulation (CM)~\cite{Llibre_MMC} makes it possible to impose both the ac and dc terminal voltages~\cite{Staffan} as long as the MMC's internal energy is well controlled. This capability is leveraged by the so-called dual-port GFM control~\cite{dualport} that imposes both IPC ac and dc terminal voltage (i.e., AC-GFM/DC-GFM) while controlling the MMC's internal energy through the ac and dc voltage.

Crucially, when using standard IPC controls, operating hybrid ac/dc systems requires careful assignment of AC-GFM/DC-GFL, AC-GFL/DC-GFM, and potentially AC-GFL/DC-GFL controls to IPCs~\cite{GoG}. For example, for a point-to-point HVDC system, at least one IPC must use DC-GFM control to control the dc voltage. Similarly, AC-GFM functions are needed when interfacing with weak grids or grids without other AC-GFM resources (e.g., offshore wind). However, in more complex hybrid ac/dc systems assigning the control roles to IPCs becomes a challenging task and small-signal stability for a given control configuration may strongly depend on the operating point~\cite{meu,meu2}. In contrast, using AC-GFM/DC-GFM control for all IPCs enables operating hybrid ac/dc system without complex control role assignment and scheduling~\cite{dualport}. 


The focus of this work is to explain and support the simulation-based results from~\cite{dualport} by using hybrid ac/dc admittance models, eigenvalue sensitivities, and experimental case studies to clarify and quantify the underlying properties of AC-GFM/DC-GFL, AC-GFL/DC-GFL, AC-GFL/DC-GFM, and dual-port GFM control in hybrid ac/dc systems.
Focusing on the impedance/admittance modeling, most of the literature only considers either the converter ac or dc admittance. Notable exceptions ~\cite{hybrid1,hybrid2} study the ac/dc cross terms in an admittance/impedance model. Specifically, \cite{hybrid1,hybrid2} analyze AC-GFL implementations for point-to-point HVDC using the generalized Nyquist criterion~\cite{hybrid1} and a single IPC using dissipativity~\cite{hybrid2}. Moreover, AC-GFL control using UCM is considered in \cite{Pedra,UCM_jeff}. In contrast,~\cite{SG2024} analyzes frequency and dc voltage stability of hybrid ac/dc systems using Lyapunov theory and crude reduced-order models of two-level VSCs. While insightful from a system perspective, the results in~\cite{SG2024} do not shed light on important properties for power electronics (e.g., harmonic stability), do not extend to common AC-GFL and AC-GFM controls, and do not cover MMCs.

In contrast, in this work, we focus on AC-GFM and DC-GFM functions and use dissipativity of the hybrid ac/dc admittance of IPCs to (i) delineate between dual-port GFM control, AC-GFM/DC-GFL control and AC-GFL control configured as either DC-GFM or DC-GFL, and (ii) analyze their dynamic interactions across the ac and dc terminals. The results highlight that, compared to common IPC controls, dual-port GFM control renders IPCs dissipative over a much wider range of frequencies and operating points.

In addition to small-signal stability, reliable operation of hybrid ac/dc systems using common AC-GFM/DC-GFL and AC-GFL/DC-GFM control crucially depends on the operating point and power flow~\cite{meu}, this aspect has not been studied for dual-port GFM control. To address this gap in the literature, we conduct an eigenvalue sensitivity analysis of hybrid ac/dc systems with respect to the operating point. The results highlight that dual-port GFM control significantly reduces the sensitivity of IPC small-signal dynamics to operating point changes compared to common AC-GFM and AC-GFL controls.

In addition to comparing the system's performance under normal conditions, this work aims to optimize standard control configurations~\cite{meu} with dual-port GFM control~\cite{dualport} under weak grid conditions and severe contingencies (i.e., loss of large power plants and/or transmission lines). A common countermeasure to limit the spread of contingencies is intentional islanding~\cite{ENTOSE_gfm, REE}, which, in hybrid ac/dc systems, can be carried out both at the IPC ac and the dc terminal. To this end, we investigate the dynamic response of a down-scaled laboratory system representing a point-to-point HVDC system for a wide range of control configurations and IPCs interconnected to a grid with low short circuit ratio (SCR), ac islanding, and dc islanding. The results demonstrate that dual-port GFM exhibits an improved dynamic response to severe contingencies.

Overall, the main contributions of this paper are (i) analysis and comparison of AC-GFM/DC-GFM and standard AC-GFM and AC-GFL IPC controls using hybrid ac/dc admittance models to clarify the resulting dynamic interactions between the IPC ac and dc terminals, (ii) using dissipativity to explain and quantify the properties of AC-GFM/DC-GFL, AC-GFL/DC-GFM, and the dual-port GFM controls, (iii) a systematic numerical and experimental comparison of hybrid ac/dc power system dynamics using the aforementioned IPC controls under severe contingencies.

This manuscript is organized as follows. Sec.~\ref{sec:roles} briefly reviews standard IPC control algorithms as well as dual-port GFM (i.e., AC-GFM/DC-GFM) control and discusses their main conceptual differences. Sec.~\ref{sec:sec3} uses hybrid ac/dc admittance models to analyze and quantify the dynamic interactions between the IPC ac and dc terminals and uses dissipativity to explain why hybrid ac/dc systems are less prone to instability when dual-port GFM controls are used. Sec.~\ref{sec:sec4} studies the small-signal dynamics of hybrid ac/dc systems and analyzes the sensitivity of the small-signal dynamics with respect to the operating point (i.e., power flow). Moreover, the response to severe transients is studied both in simulation and using a down-scaled laboratory system of a point-to-point HVDC system. Finally, Sec.\ref{sec:conclusions} provides the conclusions. 
 
\section{Interconnecting Power Converters Controls}\label{sec:roles}
From a system-level perspective, the role of an IPC depends on its control law. In general, on the ac terminal, IPC controls can be broadly categorized into grid-forming (GFM) and grid-following (GFL) controls. We emphasize that the precise definition of these roles is subject of considerable debate and resolving this challenge is beyond the scope of this manuscript. However, as this classification is important in this work, we will briefly review commonly accepted functions and properties of GFM and GFL control (please see, e.g.,~\cite{ENTOSE_gfm} for further specific characteristics required for GFM control).

\subsection{Review of common AC-GFL and AC-GFM controls}
\begin{figure}
    \centering    \includegraphics{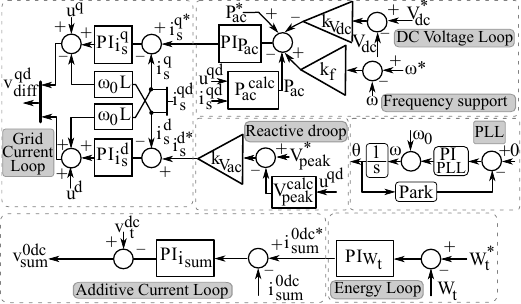}
    \caption{Standard ac grid-following control (horizontal and vertical energy balancing control is omitted for simplicity).}
    \label{fig:DC_control_scheme}
\end{figure}

AC-GFL commonly describes controls that inject/absorb active and reactive power or current while maintaining alignment with the ac grid voltage at the IPC's point of connection, usually using a phase-locked-loop (PLL)~\cite{PLL1}. Several common IPC controls found in the literature belong to this category~\cite{Vdc1, Vdc2, Vdc3, Prieto}. The most widely considered AC-GFL control structure for IPCs is depicted in Fig.~\ref{fig:DC_control_scheme} and uses a PLL to align the IPC control reference frame with the ac grid voltage. Using this control with $k_{V_{\text{dc}}} = 0$, the IPC tracks the active power reference $P_{\text{ac}}^*$ on its ac terminal and controls the internal energy $W_t$ through the dc terminal. In contrast, when using $k_{V_{\text{dc}}} > 0$, the active power exchanged with the ac grid results from a trade-off between dc voltage droop and ac frequency droop. 

\begin{figure}
    \centering
\includegraphics{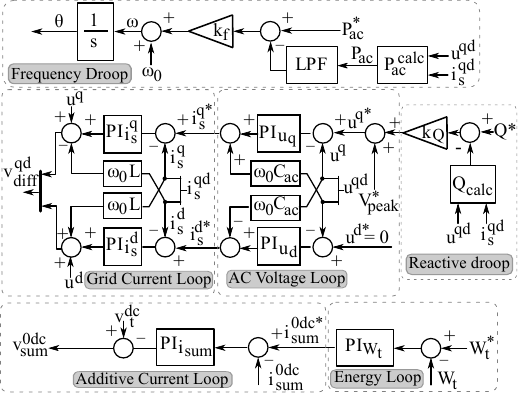}
    \caption{Standard ac grid-forming control (horizontal and vertical energy balancing control is omitted for simplicity).}
    \label{fig:gfm_scheme}
\end{figure}

In contrast, AC-GFM commonly describes controls that impose a stable ac voltage (i.e., magnitude and frequency) at the IPC ac terminal and implement $P-f$ and $Q-V$ droop using measurements of the IPC active and reactive power injection. Numerous controls documented in the literature belong to this category~\cite{GF_outer1, GF_outer2, power-frequency, pf_1, pf_2}. A prototypical implementation is depicted in Fig.~\ref{fig:gfm_scheme} that controls the voltage phase angle through a power-frequency droop. This control regulates the ac voltage using an outer ac voltage loop and inner ac current loop both containing (optional) decoupling terms.

Whether or not decoupling terms (i.e., $\omega_0 C_{\text{ac}} U_q$ and $\omega_0 C_{\text{ac}} U_d$) are used in the ac voltage loop depends on (i) the presence of a physical capacitor as part of the converter filter, (ii) the point at which the IPC voltage control is established (i.e. IPC or grid terminal of the transformer), and (iii) consideration of a potential network equivalent line capacitance at the IPC transformer terminal. IPCs for HVDC applications typically use an inductive filter without capacitor, therefore when, controlling the ac voltage (i.e., phase angle and magnitude) at the converter terminals of the transformer, $C_{\text{ac}}\!=\!0$ is used in the control. Otherwise, if the AC-GFM reference voltage is tracked at the grid terminal of the IPC transformer (see, e.g.,~\cite{CGHVDC,CCC+2022}), the equivalent transmission line capacitance $C_{\text{ac}}$~\cite{CGHVDCnet} can be used in the control to improve performance. If the equivalent shunt capacitance of the line is unknown or negligible, $C_{\text{ac}}\!=\!0$ is typically used in the control.

While the aforementioned control functions and features are generally well-defined for the ac terminal, no standard terminology exists for the dc terminals~\cite{Shinoda}. However, in emerging MT-HVDC grids, control functions also need to be assigned to the converter dc terminal (see, e.g., ~\cite{meu,meu2} and the references therein). A first attempt was made in~\cite{Shinoda}, where a general comparison between dc voltage droop control and AC-GFL and AC-GFM control is presented. Notably, while the internal energy is controlled through the dc terminal, the AC-GFL control in Fig.~\ref{fig:DC_control_scheme} may exhibit DC-GFM and DC-GFL features depending on the choice of $k_{V_{\text{dc}}}$. In particular, when $k_{V_{\text{dc}}}>0$, the IPC ac power explicitly controls the dc voltage and the internal energy is controlled through the voltage $v_{\text{sum}}^{\text{dc}0}$ applied by the converter on its dc terminal). Broadly speaking, this control may be classified as DC-GFM control. In contrast, if $k_{V_{\text{dc}}}=0$, the IPC does not control the dc voltage and may be classified as DC-GFL. Likewise, the AC-GFM control in Fig.~\ref{fig:gfm_scheme} requires a stable dc voltage and may therefore be classified as DC-GFL.

\subsection{Dual-Port GFM Control}
\begin{figure}
     \centering
     \includegraphics[width=\columnwidth]{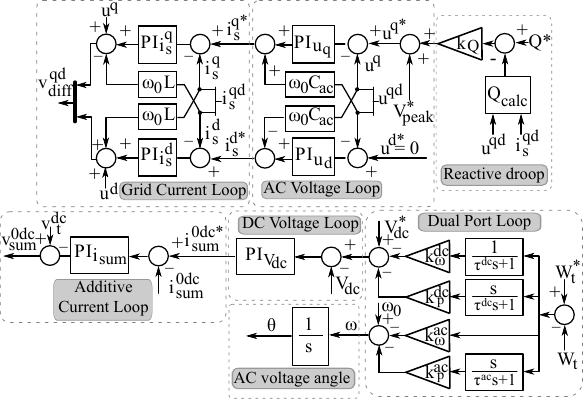}
     \caption{Dual-port GFM control (horizontal and vertical energy balancing control is omitted for simplicity).}
     \label{fig:DP_scheme}
 \end{figure}

The literature on simultaneous AC-GFM and DC-GFM control of IPCs is nascent. This paper attempts to fill this gap by focusing on the controller depicted in Fig.~\ref{fig:DP_scheme} first proposed in~\cite{dualport}. The key idea behind this control is to regulate the internal energy ($W_t$) through the voltages imposed by the IPC on its ac and dc terminals. This core functionality is implemented in the "Dual Port Loop" highlighted in Fig.~\ref{fig:DP_scheme} that consists of proportional-derivative $W_t$-$\omega$ and $W_t$-$V_{dc}$ droop. The proportional energy droop stabilizes the internal energy while the derivative energy droop provides active damping.

Dual-port GFM control has some advantages compared to the standard controls previously presented. Firstly, in the case of the standard controls, the IPC’s ac and dc terminal controllers operate independently and assume that a GFM resource consistently provides the necessary power. This approach can potentially collapse the MMCs' internal energy when the ac or the dc terminal control attempts to inject more power than is available on the opposite terminal~\cite{dualport}.

Moreover, notice that in the standard control laws, a stable ac grid is needed to implement AC-GFL control (Fig.~\ref{fig:DC_control_scheme}), i.e., AC-GFL controls are not stable when there is no other grid-forming element in the ac grid. Similarly, a stable dc grid is required for DC-GFL controls (e.g., Fig.~\ref{fig:DC_control_scheme} with $k_{V_{\text{dc}}}=0$ and Fig.~\ref{fig:gfm_scheme}). The key advantage of dual-port GFM control, compared to other controls (Fig.~\ref{fig:DC_control_scheme} and Fig.~\ref{fig:gfm_scheme}) is its stability in the absence of other devices forming both ac and dc grids~\cite{dualport}.

\section{Power Electronic Systems Considerations: Comparison between Standard Controls and Dual-Port GFM Control}\label{sec:sec3}
The choice of IPCs control role has wide ranging implications for the system dynamics. This is analyzed in detail in the first part of this Section. Moreover, the small-signal stability robustness to disturbances is another important consideration. To this end, we use a dissipativity analysis to compare how susceptible the aforementioned control roles are to instability.

\subsection{Dynamic interactions between ac and dc terminals}\label{subsec:interactions}
This section focuses on the possible dynamic interactions that might occur between the ac and dc converter terminals. 

One way to study the converter interactions is through its impedance/admittance matrix. To this end, accurate modeling of the MMC’s impedance/admittance is essential, especially to capture its internal dynamics, which can present signals at the nominal frequency, double the nominal frequency, and dc. Notably, double line frequency dynamics have to be considered when using uncompensated modulation \cite{Llibre_MMC}. In this case, representing the MMC as a linear time-invariant (LTI) system in the \textit{qd0}-frame may lead to inaccurate conclusions because dynamics at multiples of the nominal frequency are not faithfully captured. Alternative modeling methods, such as the Harmonic State Space~\cite{HSS} representation, are more suitable to study IPC controls for which multiple frequency components need to be considered.

In contrast, using compensated enables a simpler analysis in the \textit{qd0}-frame  because double line frequency components do not appear \cite{Llibre_MMC}. Moreover, we emphasize that the MMC can be modeled as separate ac and dc circuits that are solely connected through the internal energy dynamics~\cite{Staffan}. This allows the MMC to be represented as three LTI systems: an ac circuit in \textit{qd0}-frame, a dc circuit, and the internal energy storage dynamics.

{In the remainder of this study, we assume that compensated modulation is used, i.e., the MMC’s impedance/admittance is calculated in both the \textit{qd0} and dc frames, accurately capturing the dynamics of the ac and dc terminals under the assumption of a three-phase balanced ac system.} The IPC's ac terminal admittance matrix {$Y_{\text{ac}}(s)$ can be obtained from}
\begin{align}
\begin{bmatrix}
        i_q \\
        i_d
    \end{bmatrix}
    =
    \underbrace{\begin{bmatrix}
        Y_{qq}(s) & Y_{qd}(s) \\
        Y_{dq}(s) & Y_{dd}(s)
    \end{bmatrix}}_{\eqqcolon Y_{\text{ac}}(s)} 
    \begin{bmatrix}
        u_q \\
        u_d
    \end{bmatrix}.
\end{align}
{Moreover,} the IPC's dc terminal admittance matrix {$Y_{\text{dc}}(s)$ can be obtained from}
\begin{align}
    i_{\text{dc}} = Y_{\text{dc}}(s){u_{\text{dc}}}.
\end{align}

{However, as discussed before,} studying only the ac and dc terminal admittances $Y_{ac}(s)$ and $Y_{dc}(s)$ is insufficient {because} it neglects the fundamental fact that ac and dc terminal dynamics are interconnected through the internal energy dynamics. {Thus, to model the cross-coupling between the ac and dc terminals, we propose to use a hybrid ac/dc admittance model. To this end,} we introduce four new terms. Notably, $Y_{dcq}(s)$ and $Y_{dcd}(s)$ map the ac dynamics to the dc {terminal}, i.e., ac terminal perturbations affect the dc terminal if and only if $Y_{dcq}(s)$ and $Y_{dcd}(s)$ are non-zero.
Conversely, $Y_{qdc}(s)$ and $Y_{ddc}(s)$ map the dc terminal dynamics to the ac terminal dynamics, i.e., dc terminal perturbations affect the ac terminal if and only if $Y_{qdc}(s)$ and $Y_{ddc}(s)$ are non-zero.
The hybrid ac/dc admittance {model \cite{hybrid1,hybrid2} is given by}
%
%
\begin{align}
    \begin{bmatrix}
        i_q \\
        i_d \\
        i_{\text{dc}}
    \end{bmatrix} = 
    \underbrace{\begin{bmatrix}
        Y_{qq}(s) & Y_{qd}(s) & Y_{q,\text{dc}}(s) \\
        Y_{dq}(s) & Y_{dd}(s) & Y_{d,\text{dc}}(s) \\
        Y_{\text{dc},q}(s) & Y_{\text{dc},d}(s) & Y_{\text{dc}}(s) 
    \end{bmatrix}}_{\eqqcolon Y_{\text{ac},\text{dc}}(s)}
    \begin{bmatrix}
        u_q \\
        u_d \\
        u_{\text{dc}}
    \end{bmatrix}.
\end{align}

{To faciliate our analysis, the} hybrid admittance matrix of an IPC with the controls discussed in Sec.~\ref{sec:roles} is calculated using a multiple-input multiple-output (MIMO) state-space model of the IPC. The IPC electric system is modeled using the approach used in~\cite{Vdc1}, which uses the  averaged  arm model~\cite{Llibre_MMC} to model the MMC submodules. {The frequency response can be obtained by evaluating the response of the MIMO state-space model at each frequency (i.e., with IPC terminal voltages as input and IPC currents as output).}

Next, the hybrid admittance is applied to a realistic case of an MMC converter using the controllers presented before. The parameters used are defined in \ref{sec:appendix1}. Different operation points are considered to assess the impact of the dispatch on the admittance curves.

\subsection{Dual-Port GFM control}\label{subsec:dpgfmadm}
\begin{figure}[b!!!]
    \centering
    \includegraphics[width=1\columnwidth]{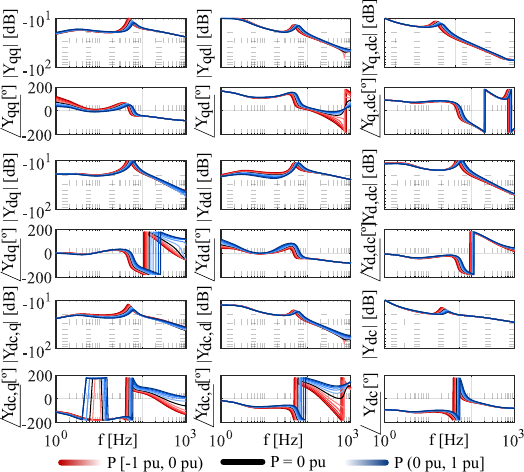}
    \caption{$Y_{\text{ac}/\text{dc}}$ for dual-port {GFM} control and different power injections}\label{fig:Y_DP}
\end{figure}

\begin{figure}
    \centering    \includegraphics[width=0.8\columnwidth]{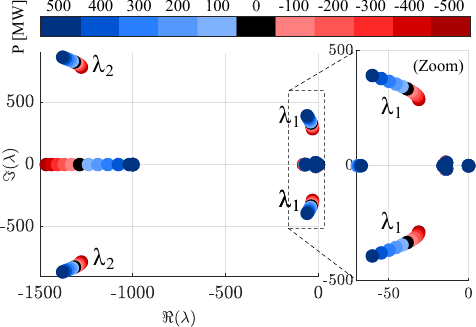}
    \caption{Eigenvalues sensitivity analysis of an IPC using dual-port {GFM} control when the IPC operation point changes.}
    \label{fig:eig_dualport}
\end{figure}

\begin{figure}
    \centering
    \includegraphics[width=0.95\columnwidth]{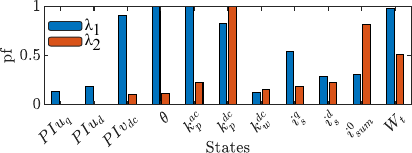}
    \caption{Participation factors of {representative eigenvalues $\lambda_1$ and $\lambda_2$} for an IPC using dual-port {GFM} control}
    \label{fig:pf_dualport}
\end{figure}

The hybrid admittance for the dual-port GFM control is depicted in Fig.~\ref{fig:Y_DP}. 
{It can be seen that dual-port GFM control induces coupling between the ac and dc terminals by controlling} the energy transfer (i) from the dc to the ac terminal through the admittances $Y_{q,dc}(s)$ and $Y_{d,dc}(s)$, and (ii) from the ac to the dc terminal through the admittances $Y_{dc,q}(s)$ and $Y_{dc,d}(s)$. {Notably,} the admittance of the IPC using {dual-port GFM} control {is largely insensitive to the operating point. The exception is the small peak at approximately $100$~Hz present in all the admittance components and slightly changes with the operating point.} This {aspect} is further analyzed using the IPC eigenvalues and participation factors. Fig.~\ref{fig:eig_dualport} shows the eigenvalue sensitivity analysis for a single IPC with dual-port {GFM} control. {The eigenvalues can be broadly categorized into three groups that exhibit sensitivity to the converter operating point.} Two {groups of eigenvalues} have {non-zero} complex parts, and therefore, an oscillatory component is expected {in the response}. {Specifically,} the complex part of $\lambda_1$ moves from {$40$~Hz to $60$~Hz as the operating point changes from $P=-1$~pu to $P=1$~pu}. {Similarly,} the complex part of $\lambda_2$ moves from $200$~Hz to $220$~Hz {as the operating point changes from $P=-1$~pu to $P=1$~pu}. To {clarify the interpretation of the dynamics associated with $\lambda_1$ and $\lambda_2$}, the participation factors of $\lambda_1$ and $\lambda_2$ are shown in Fig.~\ref{fig:pf_dualport}. The states participating in {these modes correspond to the derivative energy droop control and internal energy $W_t$. Thus, the admittance in this frequency region can be adjusted by tuning the active damping through the gains $k^{ac}_p$ and $k^{dc}_p$.} 

\subsection{AC-GFM control}
Next, the admittances of the standard control approaches are studied and compared with dual-port {GFM} control.

\begin{figure}
    \centering
    \includegraphics{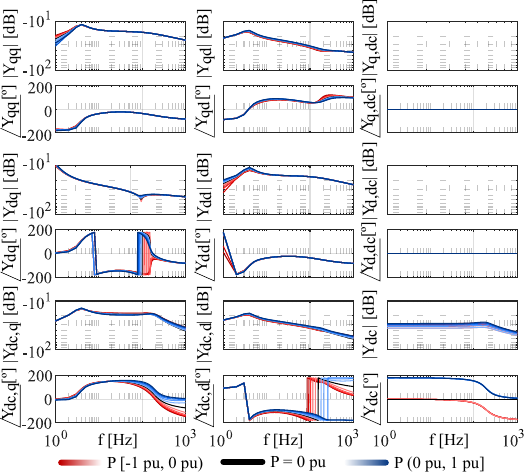}
    \caption{$Y_{\text{ac}/\text{dc}}$ for AC-GFM with different power injections}\label{fig:Y_GFM}
\end{figure}

The admittance of an IPC {implementing a} AC-GFM {controller} (Fig.~\ref{fig:gfm_scheme}) is depicted in Fig.~\ref{fig:Y_GFM}. {It can be seen that} the dc terminal is not coupled with the ac terminal {i.e.,  $Y_{q,dc}(s)$ and $Y_{d,dc}(s)$ are zero. Therefore, the perturbations on the dc terminal do not affect the ac terminal.} In this case, the converter acts as a firewall {between the dc and ac terminal}, i.e., the dynamics of the dc terminal are decoupled from the ac terminal. This is an important difference between dual-port GFM control and standard AC-GFM control.  Moreover, {considering} $Y_{dc}(s)$ it can be observed that {AC-GFM} control acts as a positive {dc} resistance ({i.e.,} $\angle{Y_{dc}(j\omega)}\in(-90^{\circ},90^{\circ})$ for $\omega$ $\in$ $(0,2\pi100)$) when the {active} power is negative ({i.e., the IPC} is absorbing power from the dc terminal). However, when the {active} power is positive ({i.e., the IPC} is injecting power {through the dc terminal}) the converter acts as a negative {dc} resistance ({i.e.,} $\angle{Y_{dc}(j\omega)}>90^\circ$ for $\omega \in (0,2\pi100)$). {In contrast, using} dual-port {GFM} control, {the IPC acts as positive dc resistance almost everywhere in} the entire frequency range ({i.e.,} $\angle{Y_{dc}(j\omega)}\in(-90^{\circ},90^{\circ})$ for $\omega \in (0,2\pi1000)$) . 

\subsection{AC-GFL control}
\begin{figure}
    \centering
    \includegraphics{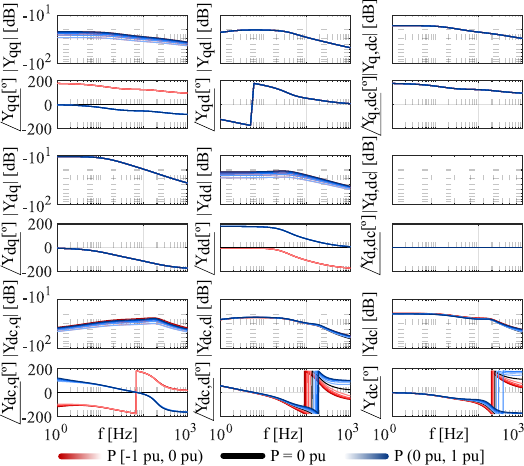}
    \caption{$Y_{\text{ac}/\text{dc}}$ AC-{GFL} with $k_{V_{\text{dc}}}>0$ with different power injections}
    \label{fig:Y_DC}
\end{figure}

The admittance of an IPC with AC-{GFL} control and $k_{V_{\text{dc}}}~>~0$ (see Fig.~\ref{fig:DC_control_scheme}) is depicted in Fig.~\ref{fig:Y_DC}. In this case, the ac and dc terminals are coupled as $Y_{q,dc}(s)$ and $Y_{dc,q}(s)$ are {non-}zero. Therefore, it has a similar behavior to dual-port control for which the perturbations from both the ac and dc terminals can cross to the opposite terminal.  Moreover, {considering} $Y_{qq}(s)$ it can be observed that this control law acts as a positive {ac} resistance ({i.e.,} $\angle{Y_{qq}(j\omega)}$ $\in(-90^{\circ},90^{\circ}))$ for $\omega \in  (0,2\pi1000)$) when the {active} power is positive ({i.e., the IPC} is absorbing power from the ac {grid}). However, when the {active} power is negative ({i.e., the IPC} is injecting power {into the ac grid}) the converter acts as a negative ac resistance ({i.e.,} $\angle{Y_{qq}(j\omega)}>90^\circ$ for $\omega \in (0,2\pi100)$). {In contrast, using} dual-port {GFM} control, the {IPC} acts as a positive {ac} resistance almost everywhere in this frequency range ({i.e.,} $\angle{Y_{qq}(j\omega)}\in(-90^{\circ},90^{\circ})$ for $\omega \in  (0,2\pi100)$). 

\begin{figure}
    \centering    \includegraphics{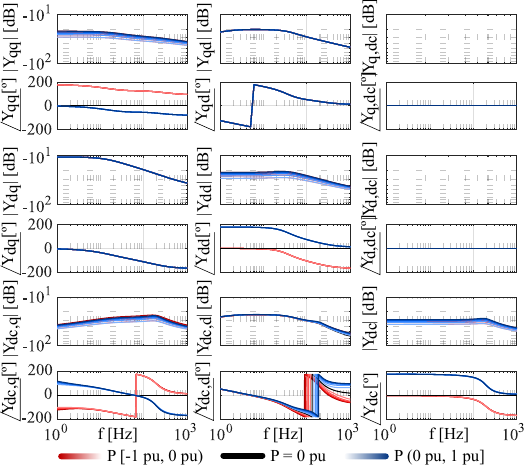}
    \caption{$Y_{\text{ac}/\text{dc}}$ AC-{GFL} with $k_{V_{\text{dc}}}=0$ with different power injections}
    \label{fig:Y_PQ}
\end{figure}

Finally, the admittance of an IPC with AC-{GFL} control and $k_{V_{\text{dc}}}=0$ (see Fig.~\ref{fig:DC_control_scheme}) is depicted in Fig.~\ref{fig:Y_PQ}. In this case, the {admittances as seen from the ac terminal (i.e., ($Y_{\text{ac}}$,$Y_{dc,q}$, and $Y_{dc,d}$)} are comparable to those obtained for AC-{GFL} control with $k_{V_{\text{dc}}}>0$, whereas the {admittances as seen} from the dc terminal {(i.e., $Y_{dc}$,$Y_{q,dc}$, and $Y_{d,dc}$)} are comparable to those obtained for AC-GFM control .

\subsection{Summary of dynamic interactions}
{The ac/dc dynamic interactions arising from different controls are summarized in Tab.~\ref{tab:table_dynamics},} where ac $\mapsto$ dc denotes that {perturbations on the ac terminal affect the dc terminal} and dc $\mapsto$ to ac denotes {perturbations on the dc terminal affect the ac terminal. Moreover, while dual-port GFM control acts as positive ac and dc resistance across a wide range of frequencies and operating points, AC-GFM control acts as negative dc resistance for a wide frequency range when the IPC is absorbing active power from the ac grid, and AC-GFL with dc voltage control (i.e., $k_{V_{\text{dc}}}>0$) acts as negative ac resistance across a wide frequency range when the IPC is injecting active power into the ac grid.}
\begin{table}
    \caption{Summary of Dynamic Interactions}
    \centering
    \begin{tabular}{ccc}
        Control Role & ac $\mapsto$ dc & dc $\mapsto$ to ac\\
        AC-GFL(Fig.~\ref{fig:DC_control_scheme} $k_{V_{\text{dc}}} = 0 $) & \checkmark & $\times$ \\
        AC-GFL(Fig.~\ref{fig:DC_control_scheme} $k_{V_{\text{dc}}}>0 $) & \checkmark & \checkmark \\
        AC-GFM (Fig.~\ref{fig:gfm_scheme}) & \checkmark & $\times$\\
        Dual-Port & \checkmark & \checkmark\\
    \end{tabular}
    \label{tab:table_dynamics}
\end{table}
{We emphasize that the results have been obtained using compensated modulation (CM)~\cite{Llibre_MMC}, which uses the arm voltage measurements to calculate the modulation indices. In contrast, using} uncompensated modulation (UCM)~\cite{UCM_3,co2}, the results {may significantly differ,} especially concerning the ac/dc {interactions}.

\subsection{{Comparative analysis using dissipativity}}
Dissipativity analysis~\cite{passivity_harnefors,passivity_saad,passivity_Zhu,passivity1} is performed to evaluate the small-signal stability and robustness of the different controls. The dissipativity properties of the IPC with various controls are evaluated using the ac admittance, dc admittance, and hybrid dc/ac admittance to study the IPC's complete dissipassivity properties~\cite{hybrid2} for various operating points. 

\begin{figure*}[htbp]
    \centering
    \includegraphics[width=\textwidth]{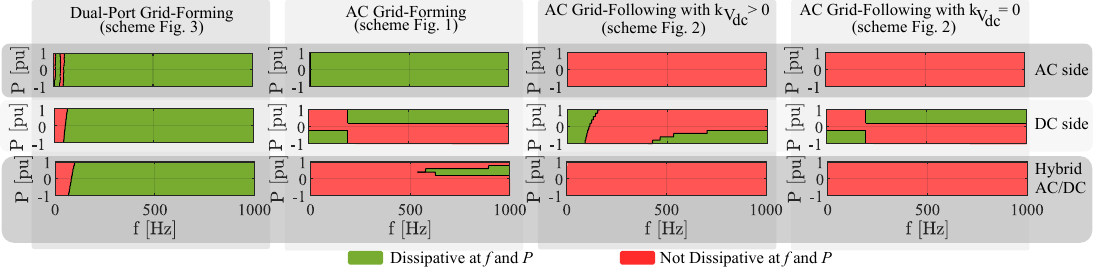}
    \caption{{Dissipativity of IPCs with different controls for frequencies from $0$~Hz to $1$~kHz and operating points from $P=-1$~pu to $P=1$~pu.}}
    \label{fig:passivity2}
\end{figure*}

Passivity of the IPC {ac terminal} can be {verified using} $Y_{\text{ac}}(s)$. The IPC {ac terminal is passive} if the smallest eigenvalue of $Y_{\text{ac}}(j\omega) + Y^{H}_{\text{ac}}(j\omega)$ is positive for all frequencies~$\omega$~\cite{passivity1}, where $Y^H$ indicates the Hermitian conjugate {of a complex square matrix $Y$}. Analogously, {the IPC} dc terminal is passive if the {smallest} eigenvalue of $Y_{\text{dc}}(j\omega) + Y^{H}_{\text{dc}}(j\omega)$ is positive for all frequencies~$\omega$.
As discussed above, analyzing {only ac and dc passivity of the} IPC {may} lead to incorrect conclusions. Therefore, {we include the ac/dc coupling terms and call the} IPC passive, if the {smallest} eigenvalue $\lambda_{min}$ of $Y_{\text{ac}/\text{dc}}(j\omega)+Y_{ac/dc}^H(j\omega)$ is positive for all frequencies $\omega$. {We emphasize that passivity constitutes a} property of the IPC across all frequencies. To  characterize the damping provided by the converter in more detail, it is insightful to evaluate aforementioned admittance matrices at specific {frequencies} and {operating points}. This concept is made precise in the following definition.
\begin{definition}{\bf{(Dissipative response at a specific frequency)}}\label{def:dissipative}
 Consider the admittance matrix $Y_{P}(j\omega)$ of an electrical subsystem at an operating point $P$ and frequency $\omega$. The electrical subsystem  is dissipative at the operating point $P$ and frequency $\omega$ if and only if the smallest eigenvalue of $Y_P(j\omega) + Y_P^H(jw)$ is larger than zero, i.e., if $Y_P(j\omega) + Y_P^H(jw)$ is positive definite.
\end{definition}
Broadly speaking, if a converter is dissipative at a given frequency and operating point, {the IPC provides damping} at this frequency and, therefore, the {system} is less prone to instability at that specific frequency and operating point. Results {for operating points from $P=-1$~pu to $P=1$~pu are depicted in Fig.~\ref{fig:passivity2}.} The frequency range is restricted to $0$~Hz to $1$~kHz to compare the dynamics of the outer loops~\cite{cigre_1khz}. The {IPC} parameters used for Fig.~\ref{fig:passivity2} are summarized in Appendix~\ref{sec:appendix1}. 

As shown in Fig.~\ref{fig:passivity2}, dual-port {GFM} control exhibits the smallest range of frequency and operating points for which the IPC is not dissipative. {Specifically,} the {IPC} ac terminal {admittance} is dissipative except for a small range at low-frequencies that are actively controlled by dual-port GFM control to ensure synchronization and stabilization of the internal energy. Similar results are obtained for the {dc terminal admittance when using dual-port GFM control}. {Next, we observe that ac/dc admittance is} dissipative for {all operating points and} frequencies higher than $100$~Hz {when using dual-port GFM control}. Notably, while the IPC with dual-port control becomes non-dissipative for lower frequencies, its {dissipativity properties are, nonetheless,} largely independent of the operating point.

Next, we note that AC-GFM control {renders the ac admittance of an IPC  dissipative for almost all frequencies and operating points}. However, {the dc admittance} is not dissipative for a wide range of frequencies and operating points. {Notably, the dc admittance is not dissipative for any frequency for the operating point $P=0$~pu widely used in analytical studies. Moreover, when the IPC is injecting active power into the ac grid ($P< 0$), the dc admittance is not dissipative for frequencies above approximately $100$~Hz. In contrast, when the IPC is injecting active power into the dc grid ($P >0$), the  dc admittance is not dissipative for frequencies below approximately $100$~Hz.} {Finally,  the hybrid ac/dc admittance is only dissipative for a small range of operating points and high frequencies when using AC-GFM control.}  This result highlights the lack of dissipativity of the IPC with AC-GFM control with respect to  interactions across the ac/dc interface. In other words, AC-GFM control provides damping to the ac grid, but only provides very limited damping to the dc grid {and for} ac/dc interactions.

The ac admittance for the remaining two AC-{GFL} controls is not dissipative at any frequency or operating point. Moreover, the dc admittance is only dissipative in a limited frequency range that strongly depends on the operating point. {Interestingly, the dc admittance of AC-GFL control without dc voltage control (i.e., $k_{V_{\text{dc}}}=0$) exactly matches the dc admittance of AC-GFM control. Finally, the hybrid ac/dc admittance is not dissipative for any frequency or operating point when using AC-{GFL} control}. {In other words, AC-GFL control provides no damping to the ac grid, only provides very limited damping to the dc grid, and provides no damping for ac/dc interactions.}

{Overall,} the passivity analysis explains and supports the observation in~\cite{dualport} that dual-port control is less prone to inducing stability in both the ac and the dc terminal. Next, to further explain and support the anecdotal simulation results in~\cite{dualport}, {system-level} stability robustness of dual-port control is compared with the standard AC-GFM and AC-{GFL} controls.

\section{{System Level Dynamics: Comparison between Standard Controls and Dual-Port GFM Control}}
\label{sec:sec4}

{In this section, the} previous small-signal robustness of dual-port {GFM} control is validated and compared against the standard controls in an ac/dc segmented system with {multiple} IPCs. {Next,} the dynamic performance of dual-port control is compared against the standard controls in an ac/dc segmented system for {various} system contingencies. Finally, dual-port {GFM} control is validated in a down-scaled laboratory system and its behaviour in a weak grid and islanded mode is compared with {standard AC-GFM and AC-GFL controls}.

\subsection{{Small-Signal Comparison}}\label{sec:small-signal}

\begin{figure}
     \centering
    \includegraphics{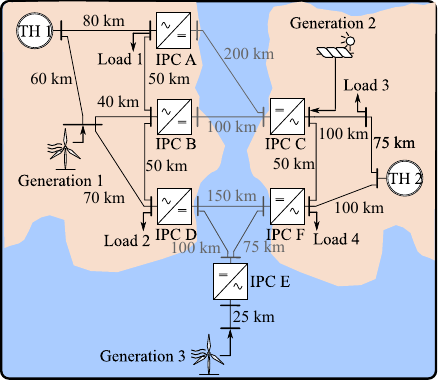}
     \caption{{Test system used for comparison of system-level responses.}}
     \label{fig:case_study}
 \end{figure}

\begin{figure}
    \centering    \includegraphics{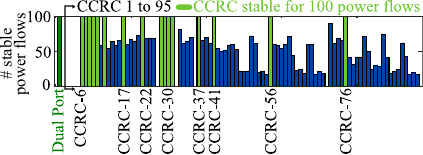}
    \caption{{Comparison of small-signal stability when using standard AC-GFM and AC-GFL control and dual-port GFM control}}    \label{fig:stability}
\end{figure}

When {studying the small-signal stability of ac/dc power systems different factors, such as assigning different control roles to the IPCs (i.e., control role configurations) and the load/generation profile and power flow need to be considered~\cite{meu}. To this end, we will first compare the impact of dual-port GFM control and standard AC-GFM and AC-GFL controls using a small-signal stability of the system shown in Fig.~\ref{fig:case_study} (please see Appendix~\ref{sec:appendix1} for a summary of system parameters). We first compare various converter control role configurations (CCRC) consisting of standard AC-GFL and AC-GFM controls (i.e., Fig.~\ref{fig:DC_control_scheme} and Fig.~\ref{fig:gfm_scheme}) to a CCRC in which all IPCs use dual-port GFM control. Specifically, the overall hybrid AC/DC power system small-signal stability is evaluated based on the eigenvalues of the linearized system dynamics at $100$ randomly generated operating points (i.e., power flow scenarios). Figure~\ref{fig:stability} illustrates the number of small-signal stable power flow scenarios (e.g., operating points) for various CCRCs using only standard AC-GFM and AC-GFL controls and a configuration that only uses dual-port GFM control. It can be seen that} dual-port control is stable for all {$100$} operating points. On the other hand, {only few CCRCs using standard controls} are stable for all {$100$ operating points.}

\begin{figure*}[b!!!]
    \centering    \includegraphics{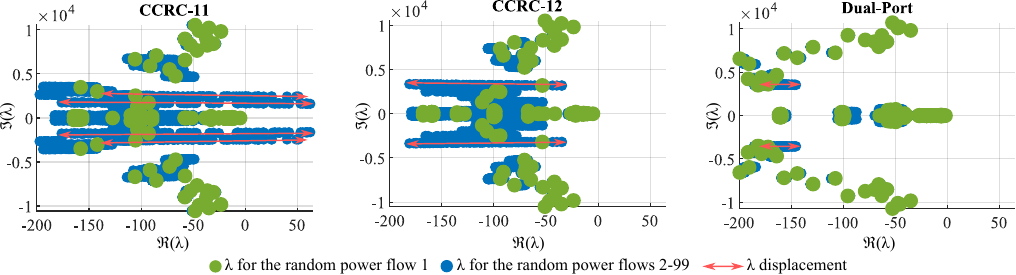}
    \caption{Comparison of eigenvalue sensitivities of standard controls and dual-port {GFM} control for several {operating points (i.e., power flow scenarios)}.}
    \label{fig:nuvol_pols}
\end{figure*}

\begin{figure}
    \centering
    \includegraphics{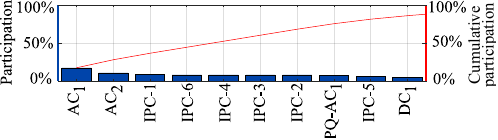}
    \caption{Pareto chart showing the {participation factors of unstable modes}. The right axis shows the cumulative participation over all the groups.}
    \label{fig:pareto}
\end{figure}

\begin{figure*}[t!!!]
    \centering
    \includegraphics[]{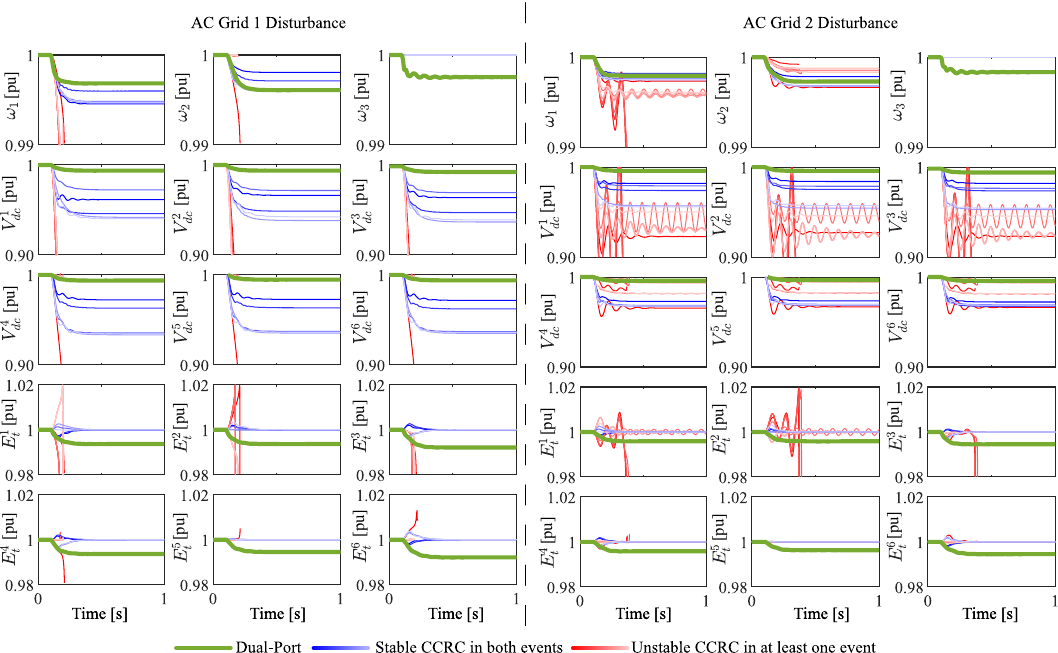}
    \caption{Comparison between dual-port {GFM} control and various {CCRCs using standard controls for} transient events}
    \label{fig:dinamica}
\end{figure*}

{To illustrate the impact of the different controls on system-level stability,} eigenvalue sensitivities of the overall system are shown in Fig.~\ref{fig:nuvol_pols} for {(i) two CCRCs that use standard controls (CCRC-11 and CCRC-12), and (ii) a configuration using dual-port GFM control.} The {assingment of controls to IPCs} for CCRC-11 and CCRC-12 is {summarized} in Tab.~\ref{tab:ccrc}. 

\begin{table}[]
    \caption{{Assignment of controllers to IPCs}}
    \label{tab:ccrc}
    \centering
    \setlength\tabcolsep{2pt}
    \begin{tabular}{cccc}
    \hline
    & \textbf{IPC-A} & \textbf{IPC-B} & \textbf{IPC-C}  \\
           \hline
       \textbf{CCRC-11}    & AC-GFM & AC-{GFL} ($k_{V_{\text{dc}}}>$0) & AC-{GFL} ($k_{V_{\text{dc}}}>$0) \\
       \textbf{CCRC-12}    & AC-GFM & AC-{GFL} ($k_{V_{\text{dc}}}>$0) & AC-{GFL} ($k_{V_{\text{dc}}}>$0)  \\
       \textbf{CCRC-22}    & AC-GFM & AC-{GFL} ($k_{V_{\text{dc}}}=$0) & AC-{GFL} ($k_{V_{\text{dc}}}>$0)  \\
       \textbf{CCRC-41}    & AC-{GFL} ($k_{V_{\text{dc}}}>$0) & AC-{GFL} ($k_{V_{\text{dc}}}>$0) & AC-GFM   \\
       \textbf{CCRC-56}    & AC-{GFL} ($k_{V_{\text{dc}}}>$0) & AC-{GFL} ($k_{V_{\text{dc}}}=$0) & AC-GFM   \\
       \textbf{CCRC-76}    & AC-{GFL} ($k_{V_{\text{dc}}}=$0) & AC-{GFL} ($k_{V_{\text{dc}}}>$0) & AC-GFM   \\
    \end{tabular}
    \begin{tabular}{cccc}
    \hline
           & \textbf{IPC-D} & \textbf{IPC-E} & \textbf{IPC-F} \\
           \hline
       \textbf{CCRC-11}    & AC-GFM & AC-{GFL} ($k_{V_{\text{dc}}}>$0) & AC-GFM \\
       \textbf{CCRC-12}    & AC-{GFL} ($k_{V_{\text{dc}}}>$0) & AC-GFM & AC-GFM \\
       \textbf{CCRC-22}    & AC-{GFL} ($k_{V_{\text{dc}}}>$0) & AC-GFM & AC-GFM  \\
       \textbf{CCRC-41}    & AC-GFM & AC-{GFL} ($k_{V_{\text{dc}}}>$0) & AC-GFM   \\
       \textbf{CCRC-56}    & AC-GFM & AC-{GFL} ($k_{V_{\text{dc}}}>$0) & AC-GFM   \\
       \textbf{CCRC-76}    & AC-GFM & AC-{GFL} ($k_{V_{\text{dc}}}>$0) & AC-GFM   \\
    \end{tabular}
\end{table}

As {shown in Fig.~\ref{fig:nuvol_pols}, the eigenvalues are highly sensitive to the operating point for CCRC-11, and the system is unstable for some operating points. In contrast, CCRC-12 results in reduced sensitivity of the eigenvalues to the operating point and stable and well-damped dynamics across all $100$ operating points. Notably, a similar dynamic behavior results for all operating points because the dominant eigenvalues are largely insensitive to the operating point. Finally, as expected from the analysis in Sec.~\ref{subsec:dpgfmadm}, we observe that using dual-port GFM control on all IPCs results in system dynamics that have the least sensitivity with respect to the operating point (see Fig.~\ref{fig:nuvol_pols}). This result highlights that dual-port GFM control minimizes the system's sensitivity against the power flow variation.}

{Based on these results,} we expect dual-port GFM control to {exhibit increased robustness} against fluctuations of load and (renewable) generation. {As predicted by the passivity analysis performed in Sec.~\ref{sec:sec3}, dual-port GFM control results in a well-damped system with few adverse interactions. In contrast, using standard AC-GFM and AC-GFL control, the causes of instability are diverse and difficult to pin-point. This observation is supported by the Pareto chart in Fig.~\ref{fig:pareto} that uses participation factors of the unstable eigenvalues to identify the main causes of instability for the unstable CCRCs. Therefore, identifying suitable CCRCs} and control parameters that are robust against {changing operating points becomes a challenging problem~\cite{meu,meu2}.}

\subsection{Dynamic Performance Comparison}
\label{sec:dynamic}
In this section, the dynamic performance {of CCRCs} that are stable for all {$100$} power flow scenarios and dual-port GFM control are compared {using two case studies}. The first {case study} considers the loss of the generation unit in ac grid 1, and the second {case study} considers the loss of the generation unit in ac grid 2. {The responses of the system to these contingencies and various CCRCs is shown in Fig.~\ref{fig:dinamica}}. The {pre-contingency} power flow used {for this study} is summarized in Appendix~\ref{sec:appendix1}. 

\begin{figure}[htbp]
    \centering
    \includegraphics{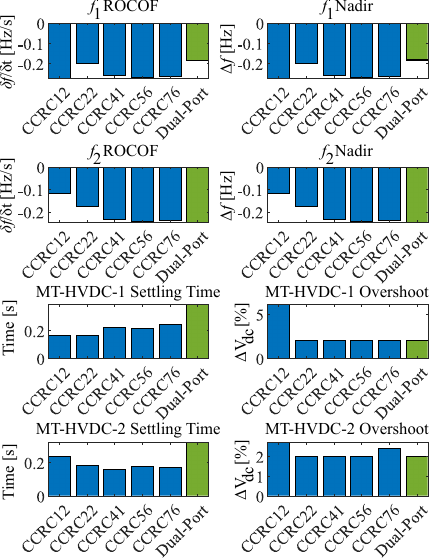}
    \caption{Quantitative study of the system dynamics {for a large contingency in ac grid 1}. Results for dual-Port {GFM} control are highlighted in green.}
    \label{fig:quantitative_ac1}
\end{figure}

\begin{figure}[htbp]
    \centering
    \includegraphics{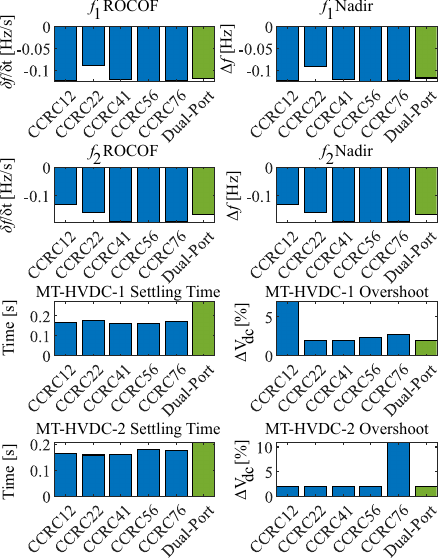}
    \caption{Quantitative study on the system dynamics {for a large contingency in ac grid 2}. Results for dual-Port {GFM} control are  highlighted in green.}
    \label{fig:quantitative_ac2}
\end{figure}

{Of the $16$ CCRCs} that were stable for all power flow scenarios considered in Sec.~\ref{sec:small-signal}, only {five CCRCs} can withstand both perturbations. On the other hand, {using dual-port GFM control for all IPCs results in a system that withstands} both perturbations (see Fig.~\ref{fig:dinamica}). The {dynamic} response of the {CCRCs} that are stable for all contingencies is {further illustrated} in Figs.~\ref{fig:quantitative_ac1}-\ref{fig:quantitative_ac2}. The rate of change of frequency (ROCOF), the frequency nadir, the mean  settling time for the dc voltages, and the mean dc voltage overshoot are used to compare the CCRCs. The main difference {is the} dc voltage settling time,  which is {consistently} slower for dual-port GFM control. In contrast, the remaining indicators {are fairly similar for dual-port GFM control and the CCRCs using standard controls. This} highlights that dual-port control can perform as well as the standard controls while {circumventing the challenging problem of identifying stabilizing CCRCs}.

Notably, dual-port {GFM} control achieves good performance {across a wide range of scenarios} because it {leverages the MMCs} internal energy to provide damping and {controls the ac frequency and dc voltage to signal power imbalance to other IPCs and generators}. As shown in Fig.~\ref{fig:dinamica}, the internal energy of the converter when using dual-port {GFM} control deviates from the nominal value {during large transients}. Therefore, it is very important to properly tune the {energy droop coefficients} $k^{dc}_{\omega}$ and $k^{ac}_{\omega}$ {to ensure} that the energy deviation does not exceed {IPC modulation limits} (see~\cite{dualport}).
Another important consideration is that dual-port {GFM} control propagates power imbalances resulting from contingencies throughout the system, confirming the analysis performed in Sec.~\ref{subsec:interactions}. In other words, a perturbation in one ac subsystem {propagates to all other subsystems and generators in a controlled manner to induce a system-wide stabilizing response}. On the other hand, some CCRCs with standard controls isolate perturbations within the perturbed ac system, as discussed in Sec.~\ref{subsec:interactions} {and, therefore, cannot benefit from the stabilizing response of generators in other subsystems.}

\subsection{{Transient Stability and Islanding} in a Down-Scaled Laboratory System}
\label{sec:transients}
{To validate the analytical results and simulations,} the controls shown in Fig.~\ref{fig:gfm_scheme}-Fig.~\ref{fig:DP_scheme} have been implemented in a down-scaled point-to-point {MMC-HVDC system using} an Imperix rapid prototyping system. {The down-scaled laboratory system} is depicted in Fig.~\ref{fig:lab_scheme} {and the parameters} summarized in Appendix~\ref{sec:appendix2}. {The} down-scaled laboratory system admits five distinct converter control role configurations that are {summarized in} Tab.~\ref{tab:kth_ccrcs} and {compared in hardware experiments.}

\begin{table}[hb!!]
    \centering
    \caption{Possible CCRC for a point-to-point {MMC-HVDC system}}
    \begin{tabular}{ccc}
    \hline
         & IPC-1 & IPC-2 \\
         \hline
      CCRC-1   & AC-{GFL} (Fig.~\ref{fig:DC_control_scheme} $k_{V_{\text{dc}}} = 0 $) & AC-{GFL} (Fig.~\ref{fig:DC_control_scheme} $k_{V_{\text{dc}}}>0 $) \\
      CCRC-2   & AC-{GFL} (Fig.~\ref{fig:DC_control_scheme} $k_{V_{\text{dc}}}>0 $) & AC-{GFL} (Fig.~\ref{fig:DC_control_scheme} $k_{V_{\text{dc}}} = 0 $) \\
      CCRC-3   &  AC-GFM (Fig.~\ref{fig:gfm_scheme}) & AC-{GFL} (Fig.~\ref{fig:DC_control_scheme} $k_{V_{\text{dc}}} = 0 $) \\
      CCRC-4 & AC-{GFL} (Fig.~\ref{fig:DC_control_scheme} $k_{V_{\text{dc}}} = 0 $)  &  AC-GFM (Fig.~\ref{fig:gfm_scheme})  \\
      CCRC-5 & Dual-Port (Fig.~\ref{fig:DP_scheme}) & Dual-Port (Fig.~\ref{fig:DP_scheme})
    \end{tabular}
    \label{tab:kth_ccrcs}
\end{table}

\begin{figure}
    \centering
\includegraphics{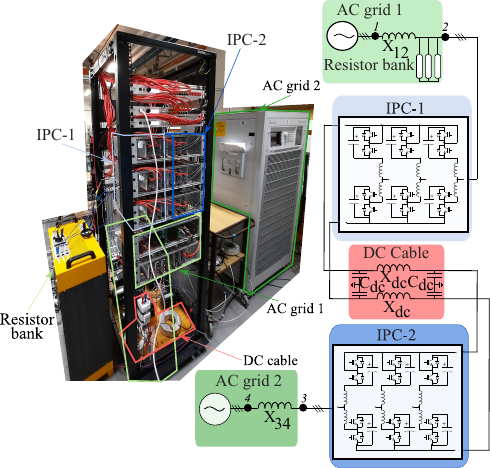}
    \caption{Down-scaled laboratory system and equivalent circuit.}    \label{fig:lab_scheme}
\end{figure}

\subsubsection{Weak Grid}
\begin{figure*}[htbp]
    \centering
    \includegraphics[]{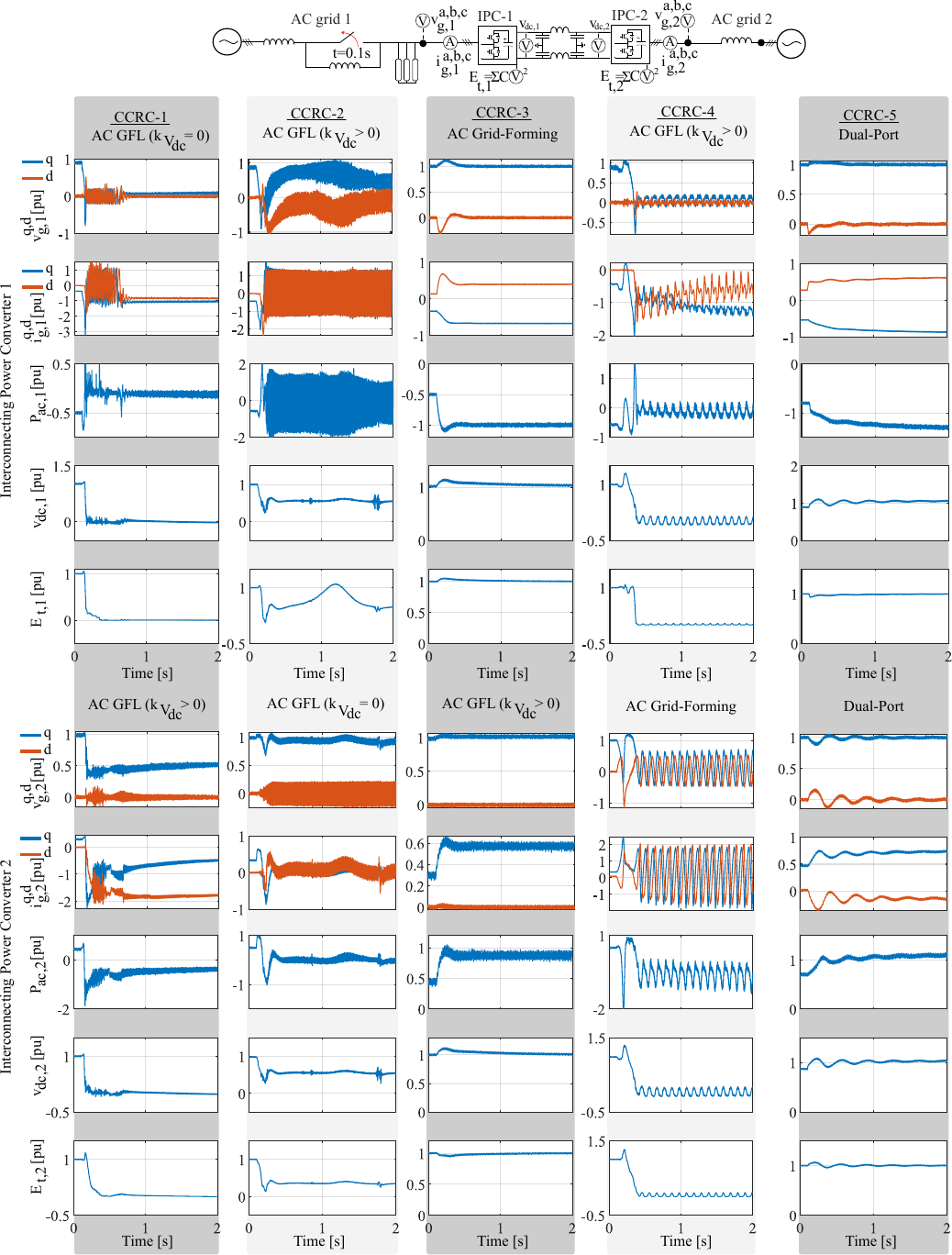}
    \caption{Comparison between dual-port {GFM} control and standard control configurations dynamics {for a reduction of SCR at} $t=0.1$~s}
    \label{fig:weak_grid}
\end{figure*}

 In the first case study, {we test a} weak ac grid {connection for IPC-1}. At time t=0.1~s, the inductance of AC-1 is {increased} to {reduce the SCR at the ac terminal of IPC-1 from} 5.5 to 2.75. The response of the {system for different CCRs is shown in Fig.~\ref{fig:weak_grid}.} Only {CCRC-3 and CCRC-5} can {maintain stable} operation {under weak grid coupling}. CCRC-3 uses AC-GFM control for IPC-1, whereas CCRC-5 uses dual-port GFM control for both IPCs. This highlights the well-known fact that {AC-GFL controls may fail under weak grid conditions while} AC-GFM controls support the operation in weak grids~\cite{GFM1, GFM3}. {Crucially, this result illustrates} that dual-port {GFM} control also {enables operation under weak grid conditions}. Note that as the system is symmetric, i.e., {identical} results would be obtained under weak grid conditions in ac grid 2. However, in this scenario, the {CCRCs} that {enable operation under weak grid conditions} would be CCRC-5, and CCRC-4 which {mirrors CCRC-3}. Note that the CCRC that uses dual-port GFM control will be the only configuration that can interconnect two weak ac grids, as all the other configurations need at least one IPC which implements a AC-{GFL} control (Fig.~\ref{fig:DC_control_scheme} with $k_{V_{\text{dc}}}>0$) to control the dc voltage. However, {the results in Fig.~\ref{fig:weak_grid} demonstrate} that AC-{GFL} control (i.e., Fig.~\ref{fig:DC_control_scheme}) is {unable to operate under weak grid coupling}.

\subsubsection{{AC islanding of IPC-1}}
\begin{figure*}[htbp]
        \centering
        \includegraphics[]{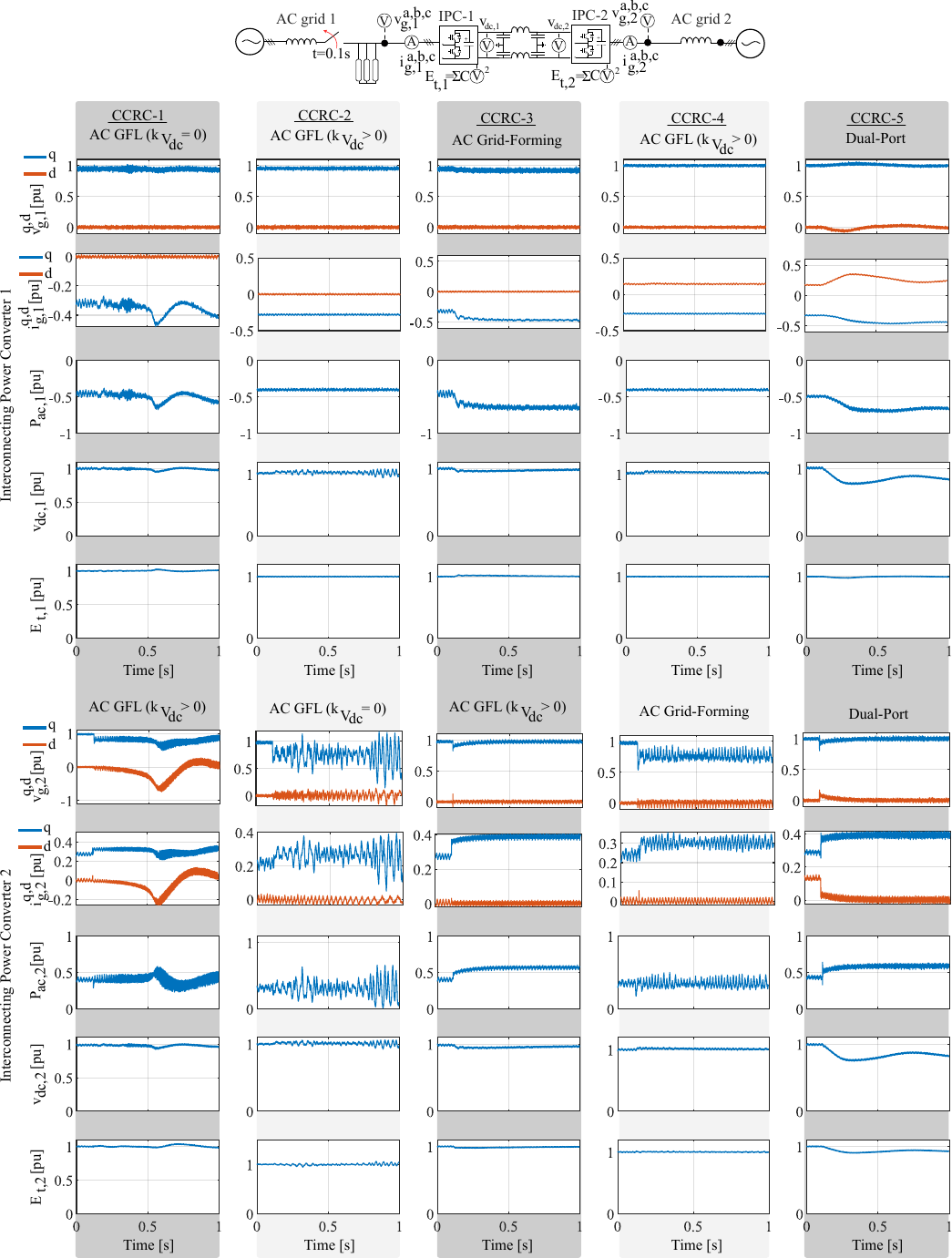}
        \caption{Comparison between dual-port {GFM} control and standard control configurations when islanding {the ac terminal of} IPC-1 at $t=0.1$~s}
        \label{fig:AC_open}
    \end{figure*}

 In this second case study, {we disconnect the ac source in ac grid 1. In other words}, at time $t=0.1$~s, the breaker in ac grid 1 {opens} and {the ac terminal of} IPC-1 is {only} connected to the load {with} no other element forming the ac grid. {The responses of the different {CCRCs} to this {contingency} are shown in Fig.~\ref{fig:AC_open}.} {Only CCRC-3 and CCRC-5  can maintain operation under ac islanding of IPC-1. {As discussed above, IPC-1 uses AC-GFM control in CCRC-3 and dual-port {GFM} control in CCRC-5.} While {CCRC-2 and CCRC-4} appear to be stable for IPC-1, it has to be noted that  the resistive load on the ac terminal of IPC-1 provides {load} damping. {For loads that provide less or no damping}, instability may occur. Moreover, {unacceptable voltage deviations and} oscillations {occur at the ac terminal of IPC-2} for both {CCRC-2 and CCRC-4}. This can be explained by the fact that}, using CCRC-2 and CCRC-4, IPC-1 regulates the dc voltage {but does not regulate the ac voltage or the ac frequency}. In summary, only configurations using AC-GFM control or dual-port {GFM} control for IPC-1 can operate {under islanding of the ac terminal of IPC-1}. {While it is well known that AC-GFM {control} can operate in islanding mode~\cite{GFM1, GFM3} our results demonstrate,} that dual-port {GFM} control can also operate {under islanding of the IPC ac terminal}.

    \subsubsection{{DC islanding}}
\begin{figure*}[htbp]
        \centering
        \includegraphics[width=1\textwidth]{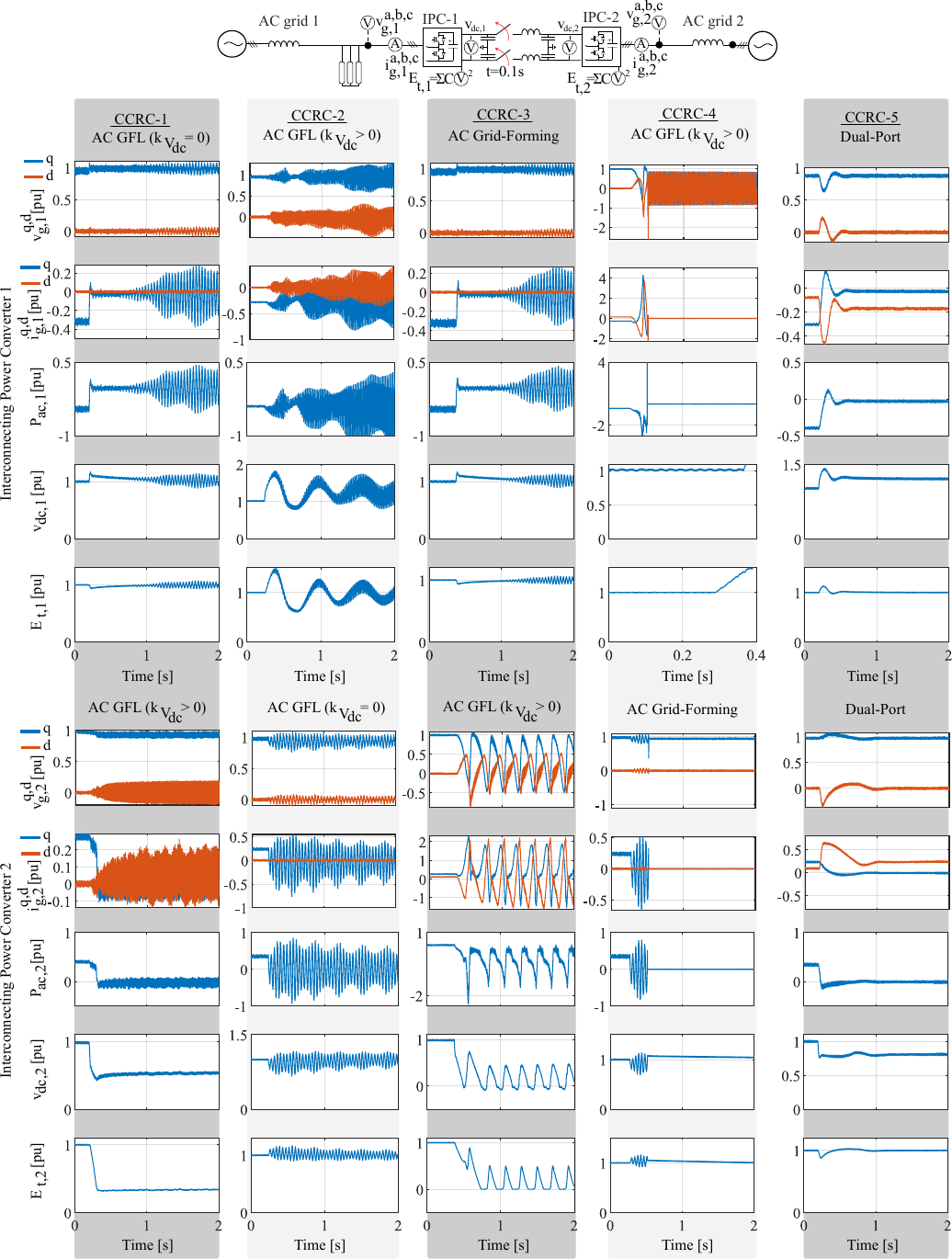}
        \caption{Comparison between dual-port {GFM} control and standard controls for {disconnecting the dc line} at $t=0.1$~s}
        \label{fig:DC_open}
    \end{figure*}

     {The final} case study {uses a} breaker to {disconnect} the dc line at  $t=0.1$~s to test the {dc grid-forming ability of the IPCs for various CCRCs}. The response of the {system to this contingency for various CCRCs} is shown in {Fig.~\ref{fig:DC_open}}. {Of the five CCRCs, only CCRC-5, i.e.,  using dual-port {GFM} control for both IPCs, maintains stable operation after this contingency. Notably, CCRC-5 is}  the only configuration {for which both IPC-1 and IPC-2 have} DC-GFM capability. Only the IPCs using {AC-GFL} control with $k_{V_{\text{dc}}}>0$ can {, in theory, maintain stability under this contingency}. However, {in practice the contingency is so severe that stability is lost for {AC-GFL} control with $k_{V_{\text{dc}}}>0$ as well.} {One of the} advantages of using dual-port {GFM} control for all IPCs is the robustness to {losing both ac and dc lines}. This can be especially interesting in MT-HVDC systems {that may still be able to operate if} some HVDC lines {are} lost when using dual-port {GFM} control. {Moreover,} if dc loads are present, these loads can be served by IPCs using dual-port {GFM control under dc islanding}.
    
\begin{figure*}[htbp]
        \centering
        \includegraphics[width=0.91\textwidth]{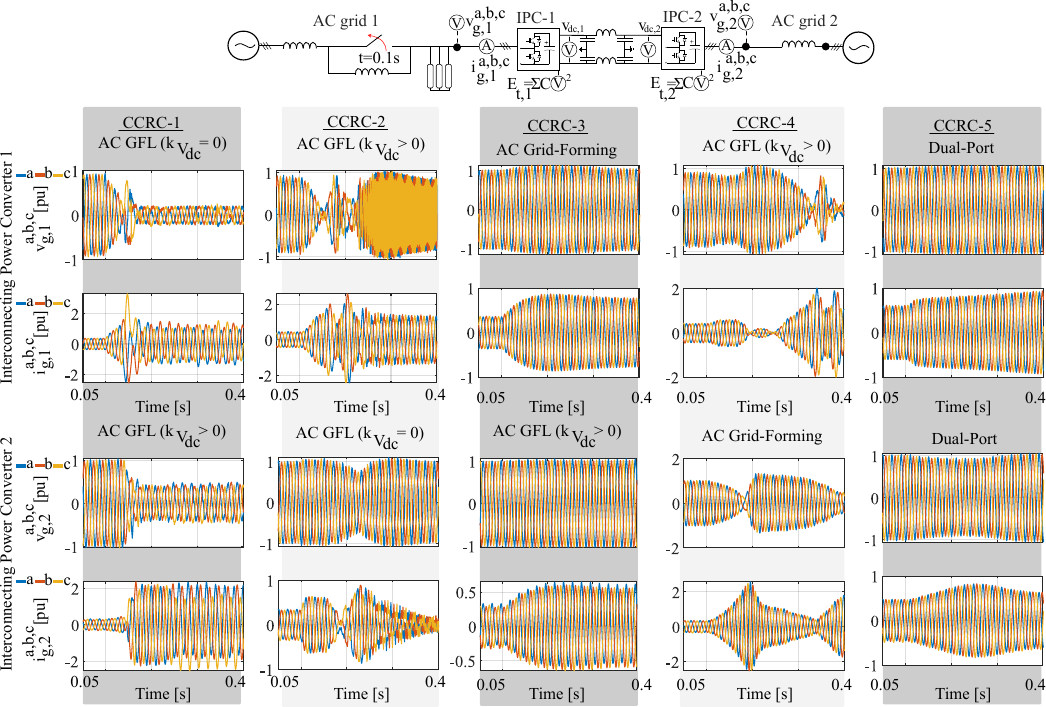}
        \caption{IPC voltage and current waveforms using dual-port GFM control and standard controls for a reduction of SCR at $t=0.1$~s}
        \label{fig:weak_grid_3p}
    \end{figure*}

\begin{figure*}[htbp]
        \centering
        \includegraphics[width=0.91\textwidth]{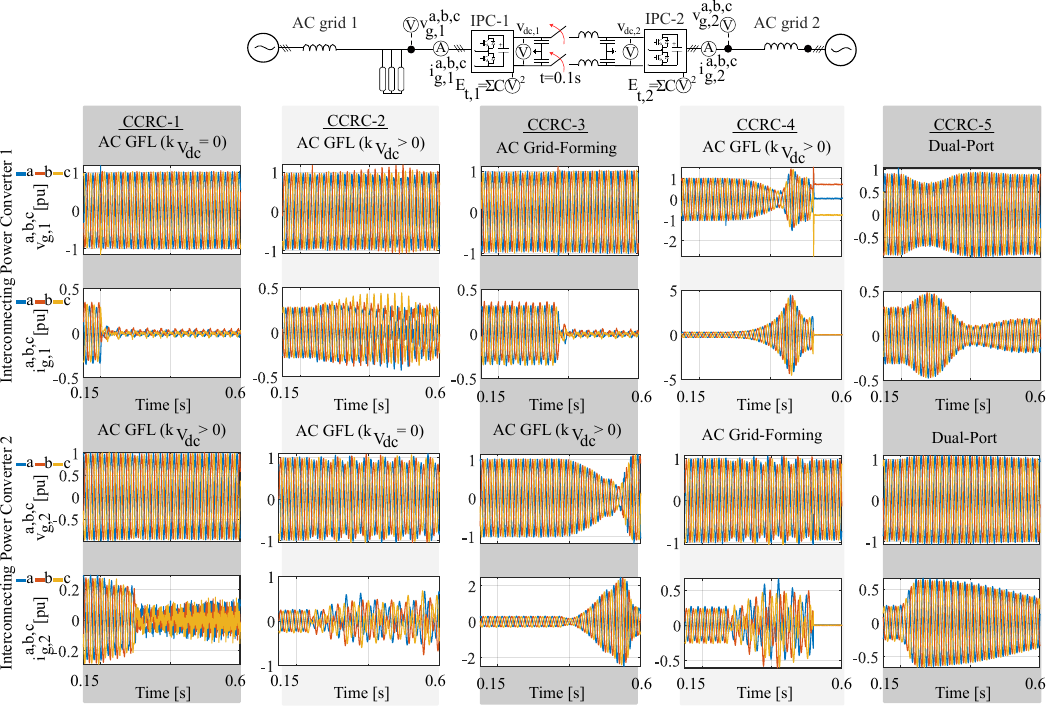}
        \caption{IPC voltage and current waveforms using dual-port GFM control and standard control under dc islanding at $t=0.1$~s}
        \label{fig:DC_open_3p}
    \end{figure*}

 For completeness, IPC voltage and current waveforms for the SCR reduction (i.e., weak grid) are shown in Fig.~\ref{fig:weak_grid_3p}, Fig.~\ref{fig:DC_open_3p} depicts IPC voltage and current waveforms for ac islanding of IPC-1, and Fig.~\ref{fig:DC_open_3p} depicts the IPC voltage and current waveforms for disconnecting the dc line.
    
    \begin{figure*}[htbp]
        \centering
        \includegraphics[width=0.91\textwidth]{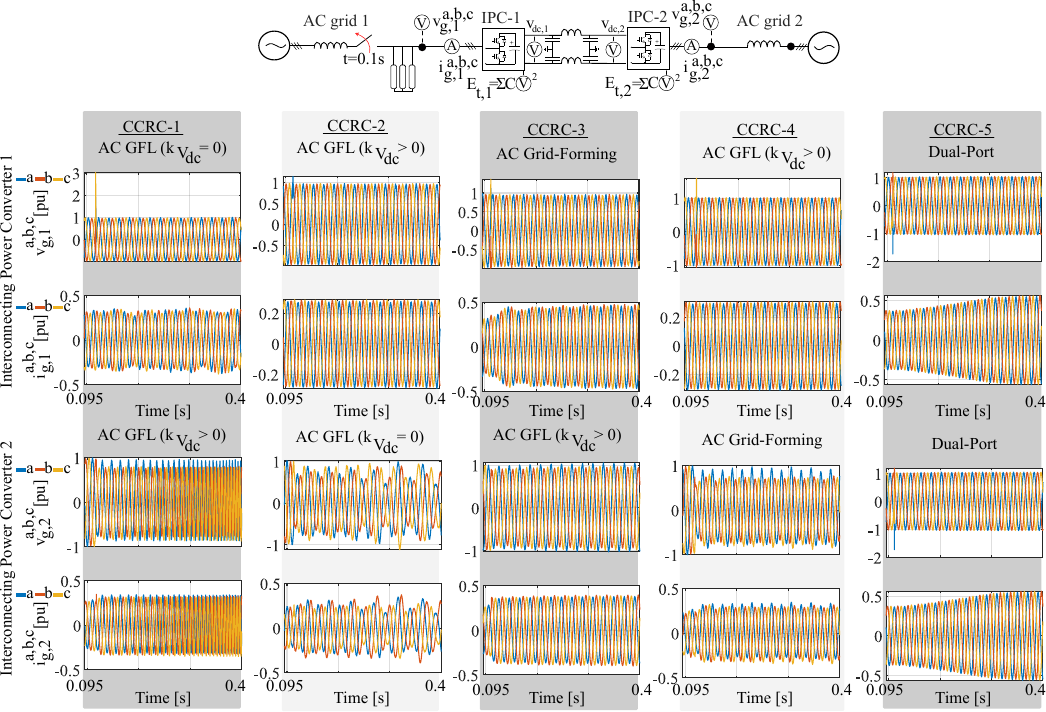}
        \caption{IPC voltage and current waveforms using dual-port GFM control and standard controls under ac islanding of IPC-1 at $t=0.1$~s}
        \label{fig:AC_open_3p}
    \end{figure*}

\section{Conclusions}\label{sec:conclusions}
In this work, {we compared} dual-port {GFM} control with {standard AC-GFM and AC-GFL control} IPCs. In particular, we investigated small-signal stability {using a hybrid ac/dc admittance approach and numerically studied} dynamic performance {and large-signal stability under severe contingencies}. Our analysis revealed that{, compared to commonly used AC-GFM and AC-GFL controls,} dual-port {GFM} control {(i) is dissipative over a much wide range of frequencies and operating points, and (ii)} is more robust to {operating point changes (i.e., power flow)}. {In our case studies,} dual-port {GFM} control also {exhibits} an {improved dynamic response to severe}  contingencies, such as the loss of generation units. Finally, {the results are illustrated and validated} using experiments performed on a down-scaled laboratory {point-to-point HVDC system}. The results {support the analytical findings and} demonstrate that dual-port {GFM} control can reliably operate in weak grids and under islanding {of} both {IPC} ac and dc terminals. Future work is envisioned to include studying the behavior of dual-port {GFM} control during faults, start-up, and black start operation.

\appendix
\subsection{Parameters used in Sec.~\ref{sec:small-signal} and Sec.~\ref{sec:dynamic}}
\label{sec:appendix1}
The IPC control gains are listed in Tab.~\ref{tab:param_control} and the following tuning guidelines have been used. The current control uses Internal Model Control to achieve a first-order closed-loop system with $\tau=1$~ms~\cite{Llibre_agusti}. The ac voltage control and parameters are documented~\cite{EnricGF}. The total energy control is designed to ensure a maximum total energy fluctuation of $10\%$~\cite{enric}. PLLs are designed to ensure tracking of the reference angle within $20$~ms~\cite{enric,Llibre_agusti}. The ac power control loop {for AC-GFL control} is tuned to achieve a first-order closed-loop response with $\tau= 10$~ms.
\begin{table}[H]
    \centering
    \caption{Control gains}
    \begin{tabular}{ccc}
        Controller & $k_p$ [pu]& $k_i$ [pu] \\
        Total energy loop & 0.019 & 0.48 \\
        AC power loop& 0.38& 177.5 \\
        AC voltage loop& 67.89 & 271.58 \\
        PLL& 1.24 & 200\\
        AC grid current loop & 0.54 & 8.65 \\
        Inner current loop & 0.73 & 11.54\\
    \end{tabular}
    \label{tab:param_control}
\end{table}
Table~\ref{tab:ipc_electric} lists {the electric parameters of the IPCs.}
\begin{table}[H]
    \caption{Interconnecting power converter electric parameters}
    \centering
    \begin{tabular}{cccc}
         Parameter & Symbol & Value & Units \\
         Rated power & $S_n$ & 500 & MVar \\
         Rated ac voltage & $U_n$ & 320 &kV rms ph-ph\\
         Rated dc voltage & $V_{dc}$ & 640 &kV \\
         Coupling impedance & $R_s+jL_s$ & 0.01 + j0.2 & pu\\
         Arm reactor impedance & $R_a+jL_a$ & 0.01 + j0.2 & pu\\
    \end{tabular}
    \label{tab:ipc_electric}
\end{table}
The control parameters {used for} dual-port GFM control are listed in Tab.~\ref{tab:dp_small_signal}.
\begin{table}[H]
    \caption{Dual-port GFM control gains}
    \centering
    \begin{tabular}[0.45\textwidth]{cccc}
         Parameter & Value & Units \\
         $k_{\omega}^{dc}$ & 1 & pu\\
         $k_{\omega}^{ac}$ & 0.496 & pu\\
         $k_{p}^{dc}$ & 0.0031 & pu\\
    \end{tabular}  
    \begin{tabular}[0.45\textwidth]{cccc}
         Parameter & Value & Units \\
         $k_{p}^{ac}$ & 0.0191 & pu\\
         $\tau^{dc}$ & 0.5 & ms\\
         $\tau^{ac}$ & 0.5 & ms\\
    \end{tabular}  
    \label{tab:dp_small_signal}
\end{table}

The AC and DC line parameters for the system in Fig.~\ref{fig:case_study} are listed in Tab.~\ref{tab:pi_parameters} and Tab.~\ref{tab:dc_lines_param}. For the ac lines, a $\pi$-section model with parameters listed in Tab.~\ref{tab:pi_parameters} is used. For the dc lines, an equivalent $\pi$-section including three parallel RL branches is used that capture the frequency-dependent cable dynamics more accurately than standard models~\cite{Cable_DC0,Cable_DCf}. The dc line parameters are listed in Tab.~\ref{tab:dc_lines_param}. The SCR for the Thevenin equivalent models of the ac grids 1 and 2 is $10$.

\begin{table}[!htbp]
\centering
\caption{$\pi$-section parameters for overhead and sub-sea lines}
\label{tab:pi_parameters}
\begin{tabular}{cccc}
        Parameter & Overhead Line & Sub-sea Line &Units \\
        $R^{a,b,c}$ & 0.08 & 0.032 & $\frac{\Omega}{km}$\\
        $L^{a,b,c}$ & 0.8 & 0.4 &$\frac{mH}{km}$\\
        $C^{a,b,c}$ & 0.012 & 0.17 &$\frac{\mu F}{km}$
\end{tabular}
\end{table}

\begin{table}[!htbp]
\centering
\caption{DC line parameters}
\label{tab:dc_lines_param}
\begin{tabular}{ccc}
        Parameter & Value & Units \\
        $R_{1}$ & 0.126 &  $\frac{\Omega}{km}$\\
        $R_{2}$ & 0.150 &  $\frac{\Omega}{km}$\\
        $R_{3}$ & 0.017 &  $\frac{\Omega}{km}$\\
        $G$ & 0.101 &  $\frac{pS}{km}$\\
\end{tabular}
\begin{tabular}{ccc}
        Parameter & Value & Units \\
        $L_{1}$ & 0.264 &  $\frac{mH}{km}$\\
        $L_{2}$ & 7.286 &  $\frac{mH}{km}$\\
        $L_{3}$ & 3.619 &  $\frac{mH}{km}$\\
        $C$ & 0.161 &  $\frac{\mu F}{km}$\\
\end{tabular}
\end{table}
Finally, the active and reactive power injection of the generators, IPCs, and loads for the case study in Sec.~\ref{sec:dynamic} are provided in Tab.~\ref{tab:powerflow}.
\begin{table}[H]
    \centering
    \caption{Active and reactive power injection for the study in Sec.~\ref{sec:dynamic}}
        \begin{tabular}[0.45\textwidth]{ccc}
         Element & P [MW]  & Q [MVar] \\
         Gen. 1 & 150 & 0\\
         Gen. 2 & 100 & 0\\
         Gen. 3 & 300 & 0\\
         Load 1 & -150 & -49.5 \\
         Load 2 & -100 & -10 \\
         Load 3 & -100 & -10\\
         Load 4 & -150 & 0\\

    \end{tabular}
        \begin{tabular}[0.45\textwidth]{ccc}
        Element & P [MW]  & Q [MVar] \\
        SG 1 & 338 & 0\\
         SG 2 & -342 & 0\\
         IPC A & -118 & 34\\
         IPC B & -238 & -0.5\\
         IPC C &  350 & -76\\
         IPC D & 126 & -59\\
         IPC E & -300 & -86 \\
         IPC F & 168 & 27\\
        \end{tabular}
    \label{tab:powerflow}
\end{table}

\subsection{Parameters of {the down-scaled hardware experiment}}\label{sec:appendix2}

The electric parameters of the down-scaled system are listed in Tab.~\ref{tab:lab_electric}.
\begin{table}[H]
    \centering
    \caption{Electric parameters of the down-scaled system}
    \begin{tabular}[0.45\textwidth]{ccc}
         Parameter & Value & Units \\
         $S_b$& 50 & W\\
         $V_b$& 14.7 & V\\
         $X_{12}$& 0.18 & pu \\
         $X_{34}$& 0.18 & pu \\
    \end{tabular}
    \begin{tabular}[0.45\textwidth]{ccc}
         Parameter & Value & Units \\
         IPC-1 arm inductance & 0.18 & pu \\
         IPC-2 arm inductance & 0.18 & pu \\
         $X_{dc}$& 0.18 & pu \\
         $C_{dc}$&  1  & mF
    \end{tabular}
    \label{tab:lab_electric}
\end{table}
The control parameters used in the IPCs for the down-scaled system are listed in Tab.~\ref{tab:lab_gen}.
\begin{table}[H]
    \centering
    \caption{Control gains used for the down-scaled system}
    \begin{tabular}{cccc}
         Controller & Proportional & Integral & Units \\
         Grid Current Loop & $1.16$ & 6.94 & pu\\
         Additive Current Loop & 11.57 & 277.66 & pu\\
         Total Energy Loop & 3.82 & 99.29 & pu \\
         Active Power Loop & 0.02 & 254.61 & pu \\
         AC Voltage Loop & 2.99 & 37.43 & pu \\
         Frequency Droop & 0.0016 & - & pu \\
         DC Voltage Loop & 0.04 & - & pu \\
         Reactive Droop & 0 & - & pu
    \end{tabular}
    \label{tab:lab_gen}
\end{table}
 The dual-port GFM control gains used in the hardware experiments are listed in Tab.~\ref{tab:lab_soa}.
\begin{table}[H]
    \centering
    \caption{Dual-port GFM control gains used for down-scaled system}
    \begin{tabular}[0.45\textwidth]{cccc}
         Parameter & Value & Units \\
         $k_{\omega}^{dc}$ & 0.6981 & pu \\
         $k_{\omega}^{ac}$ & 0.04 & pu\\
         $k_{p}^{dc}$ & 0.0233 & pu\\
    \end{tabular}
    \begin{tabular}[0.45\textwidth]{cccc}
         Parameter & Value & Units \\
         $k_{p}^{ac}$ & 0.0004 & pu\\
         $\tau^{dc}$ & 1 & ms\\
         $\tau^{ac}$ & 1 & ms\\
    \end{tabular}
    \label{tab:lab_soa}
\end{table}

\bibliographystyle{IEEEtran}
\bibliography{ieeeabrv,library}

\begin{IEEEbiography}[{\includegraphics[width=1in,height=1.25in,clip,keepaspectratio,angle=0]{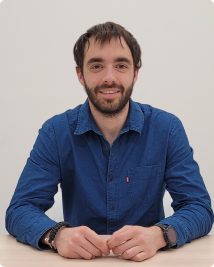}}]{Josep Arévalo-Soler} (Graduate Student Member, IEEE) received the Bachelor’s Degree in Industrial Technology Engineering and the Master’s Degree in Industrial Engineering from the School of Industrial Engineering of Barcelona, Technical University of Catalonia (UPC), Barcelona, Spain, in 2016 and 2018, respectively. He is currently in CITCEA-UPC pursuing the Ph.D. degree in electrical engineering. His current research interests include modeling, control and interaction of power electronics in AC/DC systems, with specific focus in HVDC networks.
\end{IEEEbiography}

\begin{IEEEbiography}
[{\includegraphics[width=1in,height=1.25in,clip,keepaspectratio,angle=0]{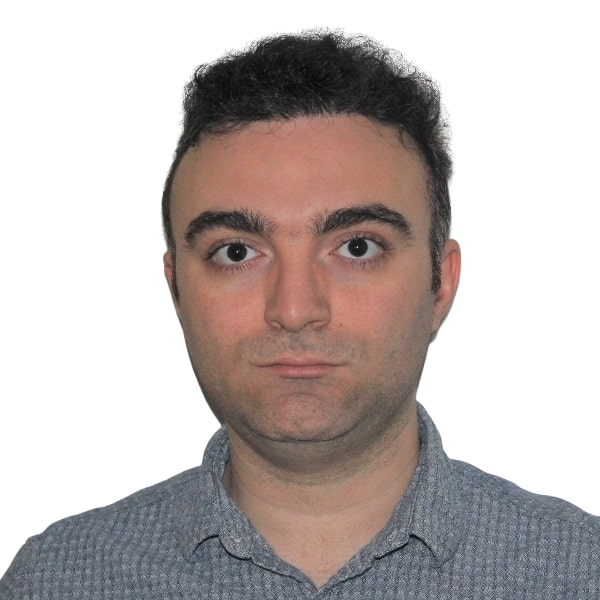}}]{Mehrdad Nahalparvari} (Graduate Student Member, IEEE) received the M.Sc. degree in power electronics from Tampere University, Tampere, Finland, in 2019. He is currently pursuing the Ph.D. degree in electrical engineering at KTH Royal Institute of Technology, Stockholm, Sweden. His research interests include modeling and control of power electronic converters.
\end{IEEEbiography}

\begin{IEEEbiography}
[{\includegraphics[width=1in,height=1.25in,clip,keepaspectratio,angle=0]{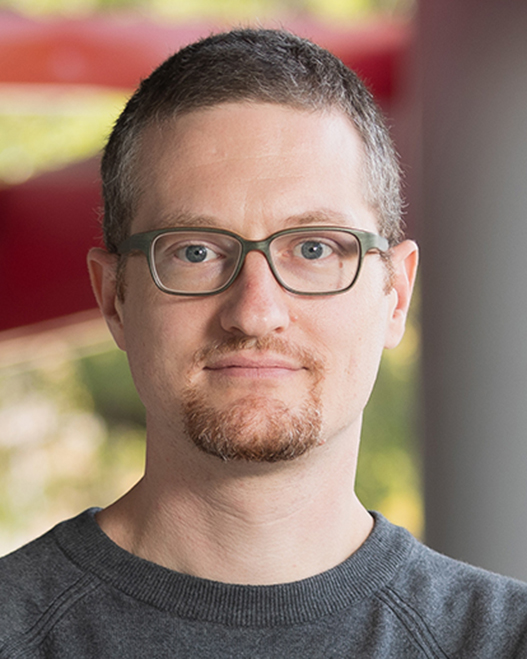}}]{Dominic Groß} (Senior Member, IEEE) received the Ph.D. degree in electrical engineering from the University of Kassel, Kassel, Germany, in 2014. He is currently a Dugald C. Jackson Assistant Professor in the Department of Electrical and Computer Engineering, University of Wisconsin-Madison, Madison, WI, USA. Prior to joining UW-Madison, he was a Postdoctoral Researcher with the Automatic Control Laboratory, ETH Zürich, Zürich, Switzerland. His research interests include distributed control and optimization of complex networked systems with applications in converter-dominated power systems and grid-forming control of power electronics.
\end{IEEEbiography}

\begin{IEEEbiography}[{\includegraphics[width=1in,height=1.25in,clip,keepaspectratio]{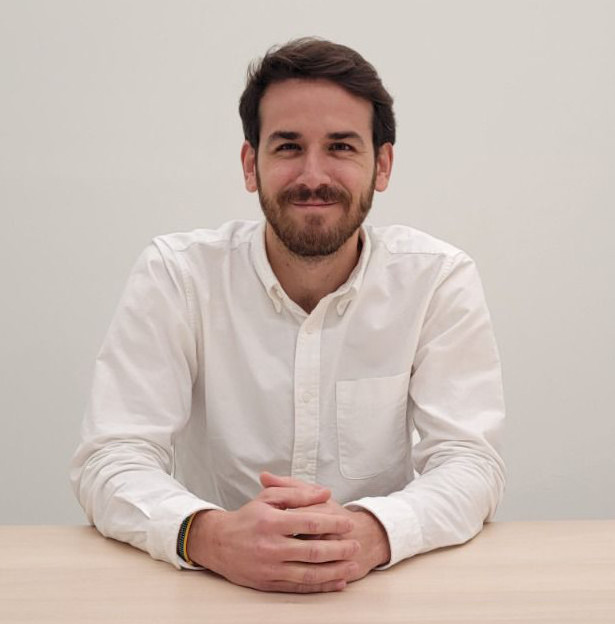}}]{Eduardo Prieto-Araujo} (S’12-M’16-SM’21) received the degree in industrial engineering from the School of Industrial Engineering of Barcelona (ETSEIB), Technical University of Catalonia (UPC), Barcelona, Spain, in 2011, and the Ph.D. degree in electrical engineering from the UPC in 2016. He joined CITCEA-UPC research group in 2010 and currently, he is a Serra Húnter Associate professor with the Electrical Engineering Department, UPC. During 2021, he was a visiting professor at the Automatic Control Laboratory, ETH Zürich. In 2022, he co-founded the start-up eRoots focused on the analysis of modern power systems. His main interests are renewable generation systems, control of power converters for HVDC applications, interaction analysis between converters, data-driven applications for power networks, and power electronics dominated power systems.
\end{IEEEbiography}

\begin{IEEEbiography}
[{\includegraphics[width=1in,height=1.25in,clip,keepaspectratio]{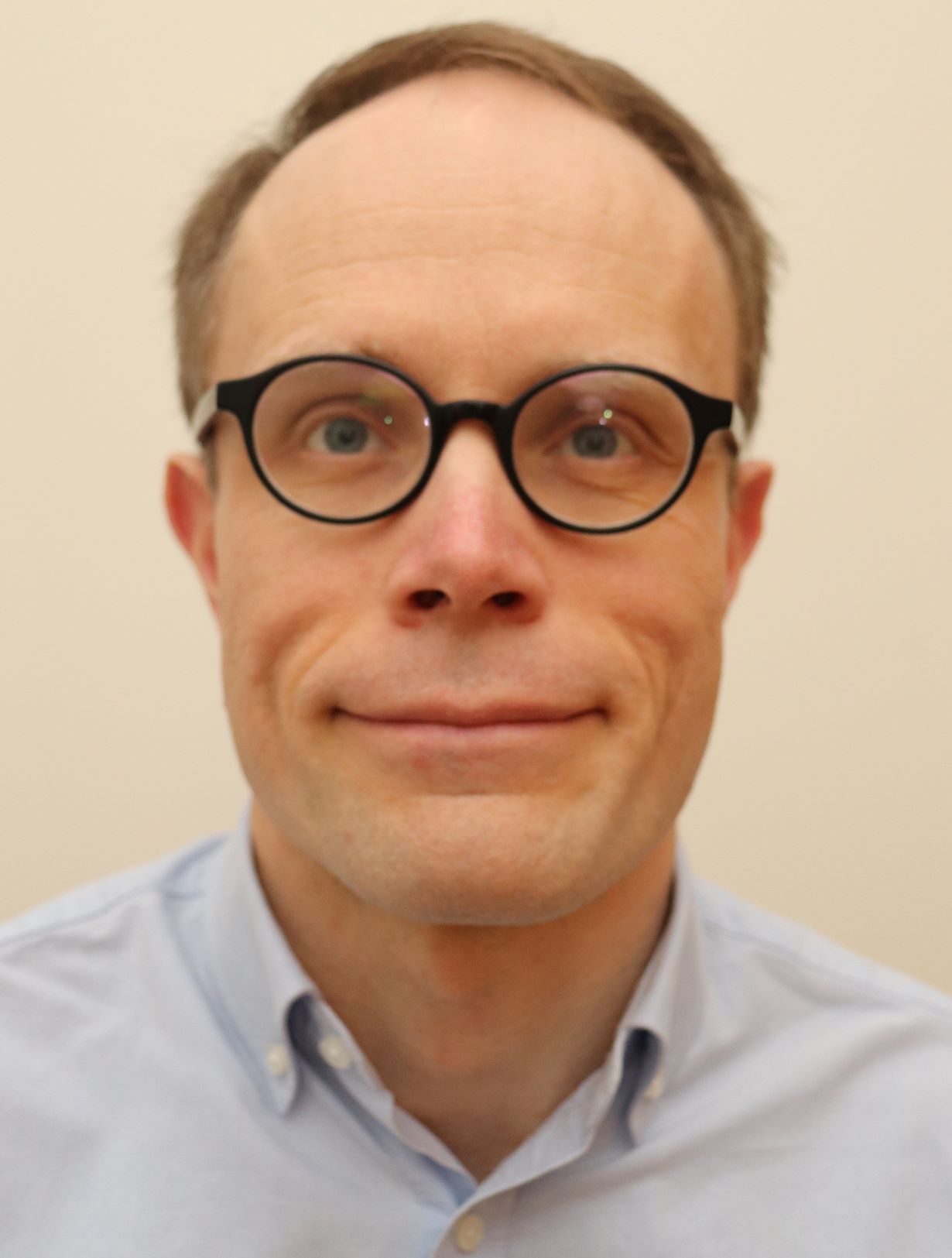}}]{Staffan Norrga} was born in Lidingö, Sweden, in 1968. He received the M.Sc. degree in applied physics from Linköping Institute of Technology, Linköping, Sweden, in 1993 and the Ph.D. degree in electrical engineering from the Royal Institute of Technology (KTH), Stockholm, Sweden, in 2005. Between 1994 and 2011, he worked as a Development Engineer at ABB in Västerås, Sweden, in various power-electronics-related areas such as railway traction systems and converters for HVDC power transmission systems. He currently holds a position as associate professor in power electronics at KTH. In 2014 he co-founded Scibreak AB which is currently part of Mitsubishi Electric. His research interests include power electronics and its applications in power grids, renewables, and electric vehicles. He is the inventor or co-inventor of more than 15 granted patents and has authored or co-authored more than 100 scientific papers published at international conferences or in journals.
\end{IEEEbiography}

\begin{IEEEbiography}[{\includegraphics[width=1in,height=1.25in,clip,keepaspectratio]{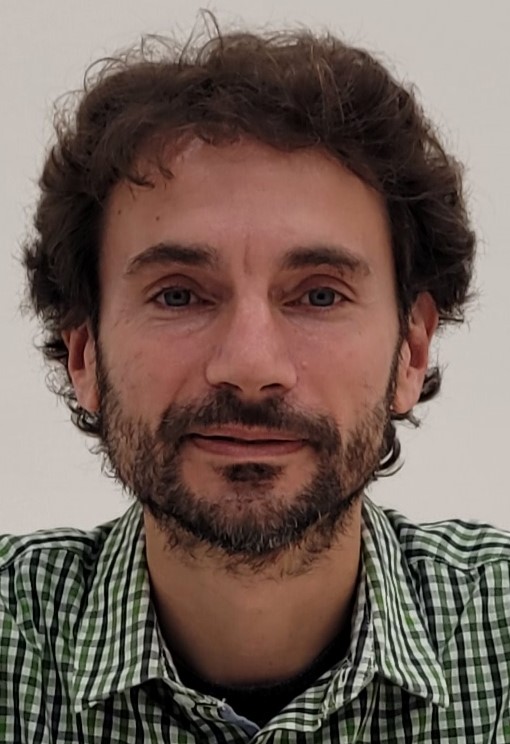}}]{Oriol Gomis-Bellmunt} (S'05-M'07-SM'12-F'21)
received the degree in industrial engineering from the School of Industrial Engineering of Barcelona (ETSEIB), Technical University of Catalonia (UPC), Barcelona, Spain, in 2001 and the Ph.D. degree in electrical engineering from the UPC in 2007. In 1999, he joined Engitrol S.L. where he worked as Project Engineer in the automation and control industry. Since 2004, he has been with the Electrical Engineering Department, UPC where he is a Professor and participates in the CITCEA-UPC Research Group. Since 2020, he is an ICREA Academia researcher. In 2022, he co-founded the start-up eRoots Analytics focused on the analysis of modern power systems. His research interests include the fields linked with power electronics, power systems and renewable energy integration in power systems.
\end{IEEEbiography}

\end{document}